\documentclass[draft]{agujournal2019}
\usepackage{rotating}
\usepackage{moreverb}
\usepackage{graphicx,overpic,natbib}
\usepackage{bm,url,color,colortbl,xcolor,soul}
\usepackage{url}
\usepackage{multirow}
\usepackage{booktabs} %for table
\usepackage{float}
\usepackage{wrapfig}
\usepackage{amsmath}
\usepackage{todonotes}
\usepackage{caption}
\usepackage[labelformat=simple]{subcaption}
\usepackage{lineno}

\newif\ifarXiv
\arXivtrue
\ifarXiv
\else
\linenumbers
\usepackage{xr}
\fi

\newcommand{\EQ}{\begin{equation}}
\newcommand{\EN}{\end{equation}}

\newcommand{\Fig}[1]{Figure~\ref{#1}}

\newcommand{\Tab}[1]{Table~\ref{#1}}

\def\nco{$N_c^{\rm obs}$ } 
\def\ncp{$N_c^{\rm predicted}$ }

\graphicspath{{./fig/}{./png/}}

\bibpunct{(}{)}{;}{a}{}{,}

\draftfalse

\journalname{Geophysical Research Letters}

\begin{document}

\title{On the Stochasticity of Aerosol-Cloud Interactions within a Data-driven Framework
}

\authors{
Xiang-Yu Li$^1$,
Hailong Wang$^1$,
TC Chakraborty$^1$,
Armin Sorooshian$^2$,
Luke D. Ziemba$^3$,
Christiane Voigt$^4$,
Kenneth Lee Thornhill$^3$
}
\affiliation{1}{Pacific Northwest National Laboratory, Richland, WA,  United States}
\affiliation{2}{Department of Chemical
and Environmental Engineering and Department of Hydrology and Atmospheric Sciences, University of Arizona, Tucson, AZ, United States}
%\affiliation{3}{University of Arizona, Department of Hydrology and Atmospheric Sciences, Tucson, AZ, United States}
\affiliation{3}{NASA Langley Research Center, Hampton, VA, United States}
\affiliation{4}{Institut f{\"u}r Physik der Atmosphäre, Deutsches Zentrum für Luft- und Raumfahrt (DLR), Oberpfaffenhofen, Germany, and Institute for Physics of the Atmosphere, Johannes Gutenberg-University Mainz, Germany}

\correspondingauthor{Xiang-Yu Li}{xiangyu.li@pnnl.gov}
\correspondingauthor{Hailong Wang}{hailong.wang@pnnl.gov}

\begin{keypoints}    
\item Three-year in-situ measurements (179 flights) provide adequate data to train and validate a random forest model (RFM) to study ACI.  
\item The RFM can successfully predict cloud droplet number concentration $N_c$ and identify importance of key predictors.  
\item Data-driven ACI in individual cases appear to be stochastic.
\end{keypoints}

\begin{abstract}
Aerosol-cloud interactions (ACI) pose the largest uncertainty for climate
projections. Among many challenges of understanding ACI, the question of
whether ACI is deterministic or stochastic has \textit{not} been explicitly
formulated and asked. Here we attempt to answer this question by predicting
cloud droplet number concentration $N_c$ from aerosol number concentration
$N_a$ and ambient conditions. We use aerosol properties, vertical velocity
fluctuation $w^\prime$, and meteorological states (temperature $T$ and water
vapor mixing ratio $q_v$) from the ACTIVATE field observations (2020 to 2022)  as
predictor variables to estimate $N_c$. We show that the climatological $N_c$
can be successfully predicted using a machine learning model despite the
strongly nonlinear and multi-scale nature of ACI. However, the
observation-trained machine learning model fails to predict $N_c$ in individual
cases while it successfully predicts $N_c$ of randomly selected data points
that cover a broad spatiotemporal scale, suggesting the stochastic nature of
ACI at fine spatiotemporal scales.
\end{abstract}

%\section*{Plain Language Summary}

\clearpage

\section{Introduction}

Atmospheric aerosols regulate Earth's energy budget directly via scattering or absorbing solar radiation and indirectly via acting as the seeds of cloud droplets, through which aerosols can alter cloud microphysical and macrophysical properties \citep{Twomey1974PollutionAlbedo, Albrecht1989AerosolsCloudiness}. Clouds modulate Earth's energy budget and water cycle (including precipitation), which, in turn, affect the sink, source, and transport processes of aerosols. This interplay between aerosols and clouds is the so-called aerosol-cloud interactions (ACI). 
ACIs remain the largest source of uncertainties for accurate
climate projections \citep{Seinfeld2016ImprovingSystem, Ghan2016ChallengesVariability, Bellouin2020BoundingChange, Bock2020QuantifyingESMValTool} due to poor
understanding of the governing processes, scarce observations, and limited computational power to resolve ACIs in numerical models. ACIs are strongly nonlinear and involve multi-scale processes with nm-sized aerosols, km-sized clouds, and hundreds of km-sized weather systems and large-scale circulations. Simulating ACI over such a wide scale range at the native scales of all the physical processes is intractable. In addition, our current physical understanding of these physical processes is incomplete. For example, the physical mechanism of the collision-coalescence of particles that is critical for precipitation and aerosol budget is still not fully understood \citep{Li2020CondensationalEnvironment, Grabowski2013GrowthEnvironment}. 

Among many challenges of understanding and quantifying ACI, the question of whether ACI is deterministic or stochastic has \textit{not} been explicitly formulated and asked. Both the condensation and the collision-coalescence processes of cloud droplets are shown to be stochastic \citep{Li2018DropletStudy, Li2022CollisionSuperdroplets} at their native spatiotemporal scales. However, these studies only focus on part of ACI pathways and their stochasticity cannot be generalized to ACI for the following reasons: 1. interactions among these cloud processes and aerosol processes are missing; 2. the mean state of ACI is not represented because of limited spatiotemporal scales.

We now face the longstanding dilemma that considering all the physical interactions across scales from nm-sized aerosols to hundreds of km-sized circulations is not feasible and focusing on part of the ACI pathways is incomplete. To tackle this dilemma and examine the stochastic nature of ACI, instead of pursuing the causality behind ACI at limited scales, here we focus on a phenomenological description of ACI based on observations. Specifically, we predict the droplet
number concentration $N_c$ from observed aerosol number concentration $N_a$,
chemical components of aerosol particles, ambient thermodynamics, and
turbulence, as these factors contribute the most to ACI metrics according to our
current scientific understanding of ACI. This is done by predicting the $N_c$ from the opportune dataset afforded by Aerosol Cloud meTeorology
Interactions oVer the western ATlantic Experiment (ACTIVATE) field measurements \citep{Sorooshian2019AerosolcloudmeteorologyCoast} using a data-driven random forest regression model (RFM) \citep{Breiman2001Random5-32.}.

For context, we first provide a brief summary of relevant ACI results in terms of $N_a$--$N_c$ relation for the ACTIVATE region. \citet{Dadashazar2021CloudFactors} showed that $N_c$ peaks in winter in contrast to $N_a$ that peaks in summer, due to stronger ACI in winter such that for a given number of aerosol particles more can activate into cloud droplets. Subsequent studies showed that seasonally, updraft velocities and turbulence (i.e., dynamics driving $N_a$ activation) are generally stronger in winter \citep{Brunke2022AircraftDevelopment}, but that microphysical/chemical attributes may be more important within a season. Also, the susceptibility of $N_c$ to $N_a$ is suspected to be stronger farther offshore of the U.S. East Coast over more remote oceans \citep{Sorooshian2019AerosolcloudmeteorologyCoast}. Processes-level modeling studies of ACI for individual cases using large-eddy simulations have shown case-dependent challenges and uncertainties in predicting $N_c$ \citep{Li2023Large-EddyInteraction, Li2023ProcessAtlantic} from available $N_a$ measurements along with information of aerosol chemical attributes and meteorological conditions, which motivates us to study ACI over a wider spatiotemporal range than individual cases. 

\section{Methods}

\subsection{Observation as training and validation data for the RFM}
\label{sec:obs-data}

To study ACIs in marine boundary-layer clouds, 179 research flights were carried out between 2020 and 2022 using a dual-aircraft approach during the ACTIVATE campaign over the Western North Atlantic Ocean (WNAO) region ($25^\circ–50^\circ$N, $60^\circ–85^\circ$W). The WNAO region is characterized by large natural and anthropogenic aerosol variability, diverse meteorological conditions, and different low-cloud regimes, which is ideal for studying ACI and for collecting unprecedented observations of aerosols, clouds, and meteorological states \citep{Corral2021AnChemistry, Painemal2021EvaluationCampaign}. The dual-aircraft approach with in-situ and remote sensing instrumentation provides coordinated, nearly simultaneous measurements of aerosol and cloud mircophysical properties as well as the ambient meteorological states. 
ACTIVATE's low-flying Falcon HU25 sampled vertical profiles by performing below cloud-base (BCB), above cloud-base (ACB), below cloud-top (BCT), and above cloud-top (ACT) flight legs, i.e., using either a stairstepping flight strategy or ``wall'' strategy involving stacked level legs \citep{Sorooshian2023Spatially-coordinatedDataset}. In-situ aerosol properties and cloud microphysical properties were measured during the BCB and ACB/BCT flight legs, respectively.
In this study, we use all the in-situ measurements of aerosol and cloud properties, turbulence, and thermodynamics during the ACTIVATE campaign.       

Aerosol
particles with diameter between $3-100\, \rm{nm}$ and larger than $100\,
\rm{nm}$ below cloud base were measured by a Scanning Mobility Particle Sizer (SMPS) and a Laser
Aerosol Spectrometer (LAS), respectively \citep{Moore2021SizingIndex}.
Mass concentrations of major aerosol chemical components (e.g., sulfate, nitrate, organics, ammonium, chloride) are
measured by an Aerodyne High Resolution Time-of-Flight Aerosol Mass
Spectrometer (HR-ToF-AMS) \citep{DeCarlo2008FastCampaign}.
The cloud microphysical properties (e.g., $N_c$) were measured by the fast cloud droplet probe (FCDP) for cloud droplets with diameter ranging from $3-50\, \mu{\rm m}$ \citep{Kirschler2022SeasonalAtlantic, Kirschler2023OverviewOcean}.
A list of data and the corresponding instruments are shown in \Tab{tab:instrument}.
Clouds are defined using the threshold of $N_c \ge 20\, \rm{cm}^{-3}$, $\text{LWC} \ge 0.02\, \rm{g\, m}^{-3}$, and effective diameter $d_{\rm eff} \ge 3.5\, \mu$m. A more comprehensive description and discussion on ACTIVATE measurements and instrument details are provided by \citet{Sorooshian2023Spatially-coordinatedDataset}. 

As we aim to predict $N_c$, all measurements are synced to FCDP-$N_c$ measurements.   
The wind speed measurements are synced to FCDP-$N_c$ measurements spatiotemporally by averaging them around each FCDP-$N_c$ data point at a given time over a window size of 20 (data points) as the sampling rate of the wind speed measurements and reported FCDP data is 20 Hz and 1 Hz, respectively.
Syncing $N_a$ and non-refractory mass concentration ($m_{\mathcal{X}}$) of aerosol chemical components to FCDP-$N_c$ measurements are more challenging because they were not strictly collocated at the native sampling frequency (i.e., aerosols sampled during BCB flight legs and cloud microphysics measured during ACB or BCT flight legs). We overcome this challenge by averaging $N_a$ and $m_{\mathcal{X}}$ over all BCB legs of each flight, assuming that aerosols below cloud base are representative of sampling region. Other than the collocation reasoning, this syncing strategy is justified by the fact that aerosol measurements (AMS, SMPS, and LAS) at each BCB flight leg only lasted for $\sim 3$ minutes for
a flight ($\sim 30$ minutes for a entire flight with $\sim 10$ BCB sampling legs; see \Fig{num_duration_BCB}). Even though this syncing strategy ignores aerosol transport during BCB-leg sampling, the statistical properties of aerosols over the measurement domain are well represented. 
The vertical velocity fluctuation $w^\prime = w - \overline{\langle w \rangle}$ (the same for $u^\prime$ and $v^\prime$) is calculated from the native 20 Hz data for each flight leg, where $\overline{\langle w \rangle}$ is obtained. 
Overall, the training and validation data for the machine learning (ML) model are based on subsets of a sample size of 69,159 with a sampling rate of 1 Hz.

Another challenge of using the in-situ aircraft measurements as the training and validation data for the ML model is to find a physical spatial scale that can represent the characteristic scales of aerosols and cloud droplets of the targeted cloud systems. This is because the aircraft performs measurements of aerosol and cloud droplet properties with a speed of about $100\, \rm{m\, s}^{-1}$. The 1 Hz data corresponds to a length scale of $100$ m, which is too small to represent typical length scales of boundary-layer cloud systems. To work around this issue, we perform a running-average of the input data and examine the $r^2$ validation score as a function of the running-average window size. The $r^2$ saturates (hits 0.99) at a window size of 20 data points (from 1 Hz data) as listed in \Tab{tab:windowSize} (the corresponding comparison of \ncp and \nco is shown in \Fig{fig:Nc_predicted_obs_20window}). Therefore, we apply a running-average window of 20 (data points) to all the 1-Hz observational data to obtain smoothed datasets for training and validating the RFM model. More importantly, this window size corresponds to a length scale of 2 km that is more representative of the length scale of cloud systems being studied here.    

There are limitations to the ACTIVATE dataset, which are generally innate to in-situ aircraft datasets, including how the sub-cloud measurements are not simultaneously collected directly underneath in-cloud measurements but are slightly offset in time and space. 
Moreover, there are instrument uncertainties (see \Tab{tab:instrument}) that potentially contribute to biases albeit state-of-the-art instruments were deployed in the ACTIVATE field campaign. Although ACTIVATE collected the largest in-situ dataset with spatially coordinated aircraft to our knowledge, statistics and spatiotemporal coverage (3 years over $25^\circ–50^\circ$N, $60^\circ–85^\circ$W) can still improve.

\subsection{Random Forest Model}

One of the objectives of this study is to predict $N_c$ from in-situ measurements of aerosol ($N_a$ and $m_{\mathcal{X}}$), turbulence ($u^\prime$, $v^\prime$, and $w^\prime$), and thermodynamics ($T$ and $q_v$), i.e., to construct a function representing 
\begin{equation}
N_c = \mathcal{G}(m_{\mathcal{X}}, N_a, w^\prime, u^\prime, v^\prime, T, q_v,
\bm{x}, \theta_z).
\end{equation}
Here $\bm{x}$ and $\theta_z$ denote the location (latitude, longitude, and altitude) and zenith angle, respectively.  
The random forest model (RFM) is chosen to achieve this because it is effective in prediction and does not overfit due to the Law of Large Numbers \citep{Breiman2001Random5-32.}. In addition, the RFM has the advantage of accurate prediction of nonlinear complex systems and easy implementation and physical interpretation of the prediction compared to other machine learning methods (e.g., neural network or deep leaning techniques) \citep{Breiman2001Random5-32.}. It is also fast due to the parallelizability in building decision trees. A random forest, as the name suggested, is an ensemble of decision trees that are a non-parametric supervised learning method used for classification and regression. The growth of each tree is governed by independent random vectors in space and time. After a numerous number of trees is generated from subsets of training data, they vote for the most popular class at input variables to make an aggregated prediction.

To determine the feature importance of predictors for $N_c$, we adopt the permutation feature importance (PFI) technique. PFI measures the contribution of each feature to a fitted model’s statistical performance on a given tabular dataset. This technique is particularly useful for nonlinear estimators, and involves randomly shuffling the values of a single feature and observing the resulting degradation of the model’s score. We note PFI \textit{does not} reflect the physical importance of a feature to a complex system but reflects how important this feature is for a particular data-driven model \citep{Breiman2001Random5-32., Pedregosa2011Scikit-learn:Python}. It is, however, important to place the PFI based on physical understanding of a complex system, to better understand what the model is really capturing, as we will discuss for the ACI studied here. 
We cross validate the permutation importance using the K-fold (10 fold) and Monte-Carlo (50 random sampling) validation method. The relative root mean square error (RRMSE) and mean normalized error (MNE) are adopted to evaluate the predictions. 

RFM has been widely used in Earth Systems and shown to be a robust tool for prediction and inference \citep{ArjunanNair2020UsingMeasurements, Chakraborty2021UsingDataset, Chen2022MachineCover, Michel2022EarlyReconstruction, Dadashazar2021CloudFactors}.
Here we limit our discussion on the application of RFM to ACI. \citet{ArjunanNair2020UsingMeasurements} showed that RFM is highly robust in predicting number concentration of cloud condensation nuclei (CCN) with the atmospheric state and composition variables as predictors from a global chemical transport model and can learn the underlying dependence of CCN on these predictors. Subsequent works further show that RFM can learn aerosol size information \citep{Nair2021MachineParticles} and can reduce uncertainties of climate models in predicting particle number concentration and radiative forcing associated with ACI \citep{Yu2022UseModels}.      
We use the RFM implemented in open-source scikit-learn \citep{Pedregosa2011Scikit-learn:Python}.

The coefficient of determination, $r^2$ score, converges at hyperparameters maximum tree depth = 30, number of of trees = 98, and test size = 0.1 (i.e., 10\% data for the validation and 90\% data for the training), which are used for all the training in this study.

\subsection{Large-eddy simulation}

\subsubsection{Numerical experiments setup}
We use the Weather Research and Forecasting (WRF) model \citep{Skamarock2019A4} in its LES configuration (WRF-LES) with doubly periodic boundary conditions \citep{Wang2009EvaluationInteractions}. The LES
domain has a lateral size of $L_x=L_y=20\, \rm{km}$ (60 km) with a grid spacing of
$dx=dy=100\, \rm{m}$ (300 m) for the cumulus (cold-air outbreak) cases and a vertical extent of $z_{\rm top} = 7\, \rm{km}$ with
153 vertical layers. Initial profiles of temperature, humidity, and horizontal wind components and the time-varying large-scale forcings and surface turbulent heat fluxes are obtained from the fifth generation of European
Centre for Medium-Range Weather Forecasts’s Integrated Forecast System (ERA5) reanalysis.
We use
the two-moment Morrison cloud microphysics scheme
\citep{Morrison2009ImpactSchemes} with prescribed aerosol size modes and hygroscopicity derived from the
ACTIVATE campaign measurements.
All simulations start at 06:00 UTC
and end at 21:00 UTC with a fixed time step of $1\, \rm{s}$. The two cold-air outbreak cases on 28 February 2020 and 1 March 2020 are characterized by a well-defined boundary layer while the two precipitating summertime cumulus cases on 02 and 07 June 2021 are characterized by strong spatiotemporal variations of cloud top height. We refer readers to \citet{Li2023Large-EddyInteraction} and \citet{Li2023ProcessAtlantic} for details of the simulations used in the present study.

\subsubsection{LES validation data for the observation-RFM emulator}

To use the observation-RFM as an emulator for the LES $N_c$ , we prepared the input data ($N_a, T, q_v, \text{and}\, w^\prime$) from LES. These predictors are chosen for the following reasons: 1. they represent our current understanding of physical processes (e.g., aerosol activation and condensation and collision-coalescence of cloud droplets) governing the ACI; 2. the RFM can successfully predict $N_c$ use these independent predictors as will be discussed in section~\ref{sec:obs-RFM}; 3. to apply the observation-RFM emulator to LES, the same variables from LES are required as input. The WRF-LES used in this study does not have prognostic aerosols, i.e., $m_\mathcal{X}$. The predicted $N_c$ from the observation-RFM emulator will be compared to the observation. This requires a point-to-point data comparison between LES and FCDP.
The same threshold values of ${\rm LWC}=0.02\, \rm{g\, m}^{-3}$, $3.5\,\mu{\rm m} \le d_{\rm eff}
\le 50\,\mu{\rm m}$, and $N_c=20\, \rm{cm}^{-3}$ as in section~\ref{sec:obs-data} are applied to LES data to define clouds.
The measurement was taken during 16:00-17:00 UTC
on 28 February 2020 and 15:00-16:00 UTC on 1 March 2020 and 19:00-20:00 UTC on both 02 and 07 June 2021.
The LES data are based on three snapshots 30 minutes apart.
We only use ACB and BCT flight legs.
The data are binned at those heights of flight legs with a
residual range of $\pm 40\, {\rm m}$ such that at least one model layer is
counted at the height of each flight leg. This also avoids double counting.
At each height, we randomly draw the same number of data points from the LES domain as in the FCDP field measurements. A single random sampling is representative as shown in \Fig{fig:r2_FCDP_LES}.

\section{Results}

\subsection{A successful data-driven prediction of $N_c$}
\label{sec:obs-RFM}
%\subsection{Determining precursor of $N_c$}

\begin{figure*}[t!]\begin{center}
  %---Nc---
    \begin{subfigure}[b]{0.32\textwidth}
          \subcaption{}
          \includegraphics[width=\textwidth]{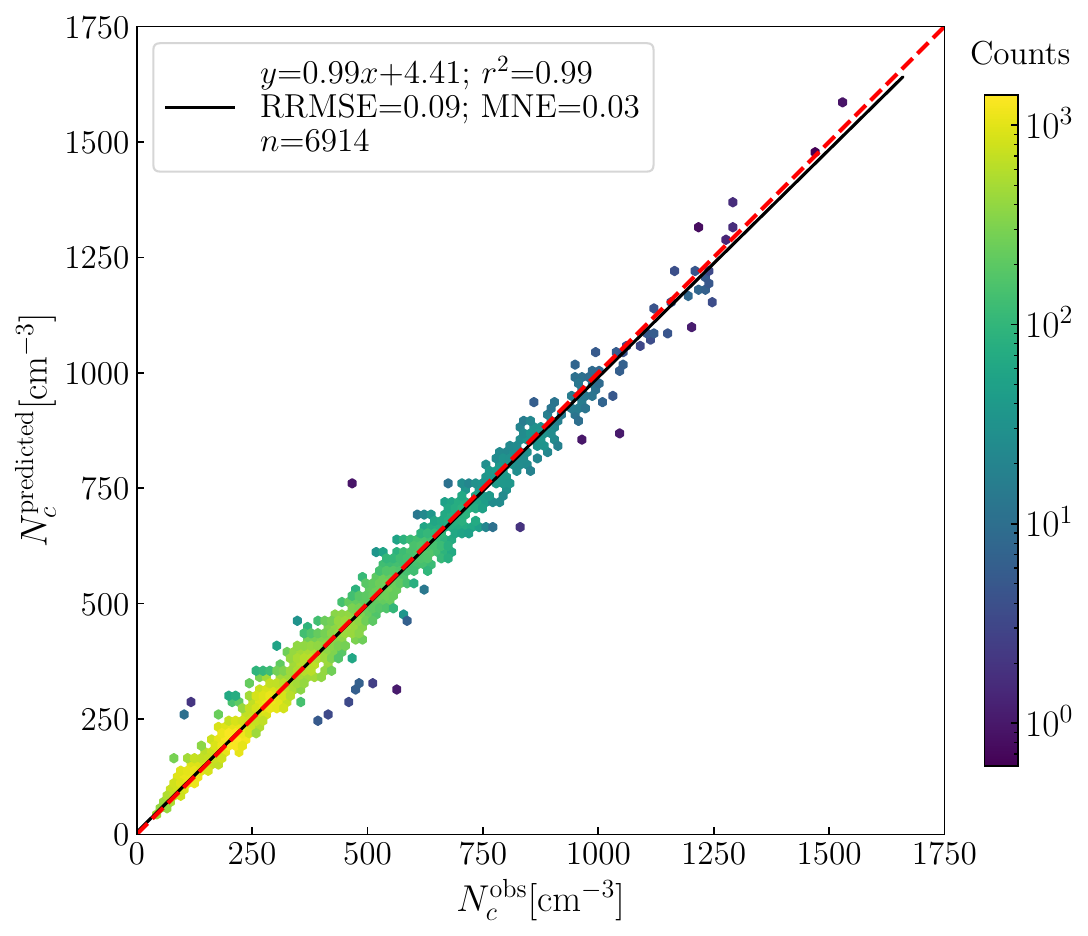}
          \label{fig:Nc_predicted_obs-a}
    \end{subfigure}
    \begin{subfigure}[b]{0.32\textwidth}
          \subcaption{}
          \includegraphics[width=\textwidth]{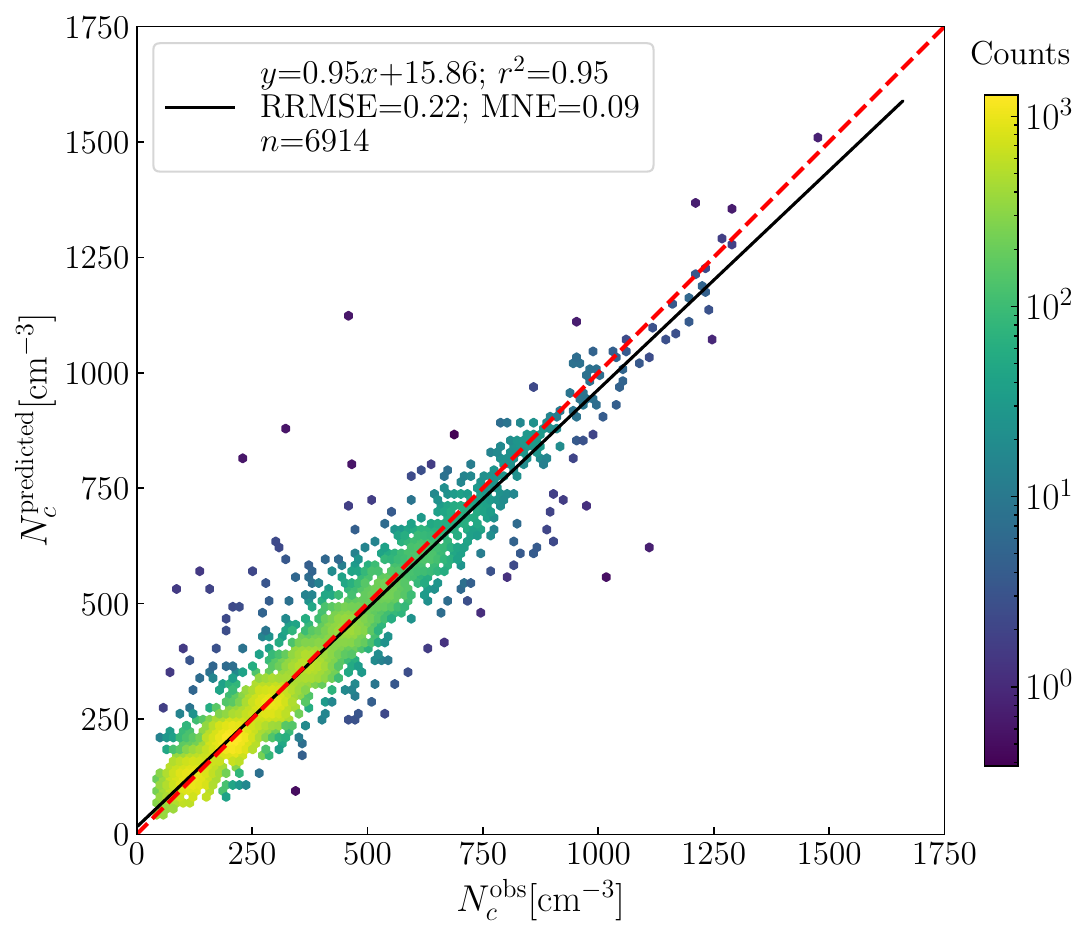}
          \label{fig:Nc_predicted_obs-d}
    \end{subfigure}
    \begin{subfigure}[b]{0.32\textwidth}
          \subcaption{}
          \includegraphics[width=\textwidth]{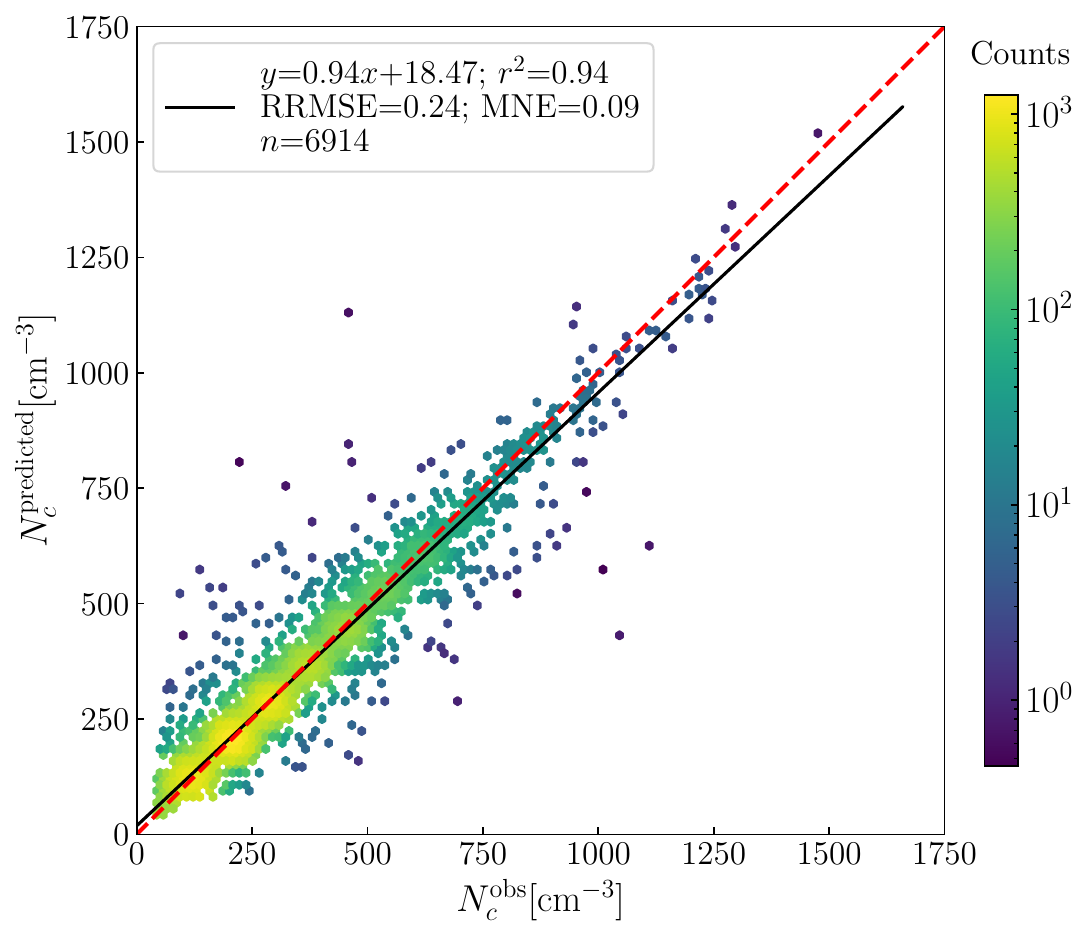}
          \label{fig:Nc_predicted_obs-g}
    \end{subfigure}
  %---K-folds---
    \begin{subfigure}[b]{0.32\textwidth}
          \subcaption{}
          \includegraphics[width=\textwidth]{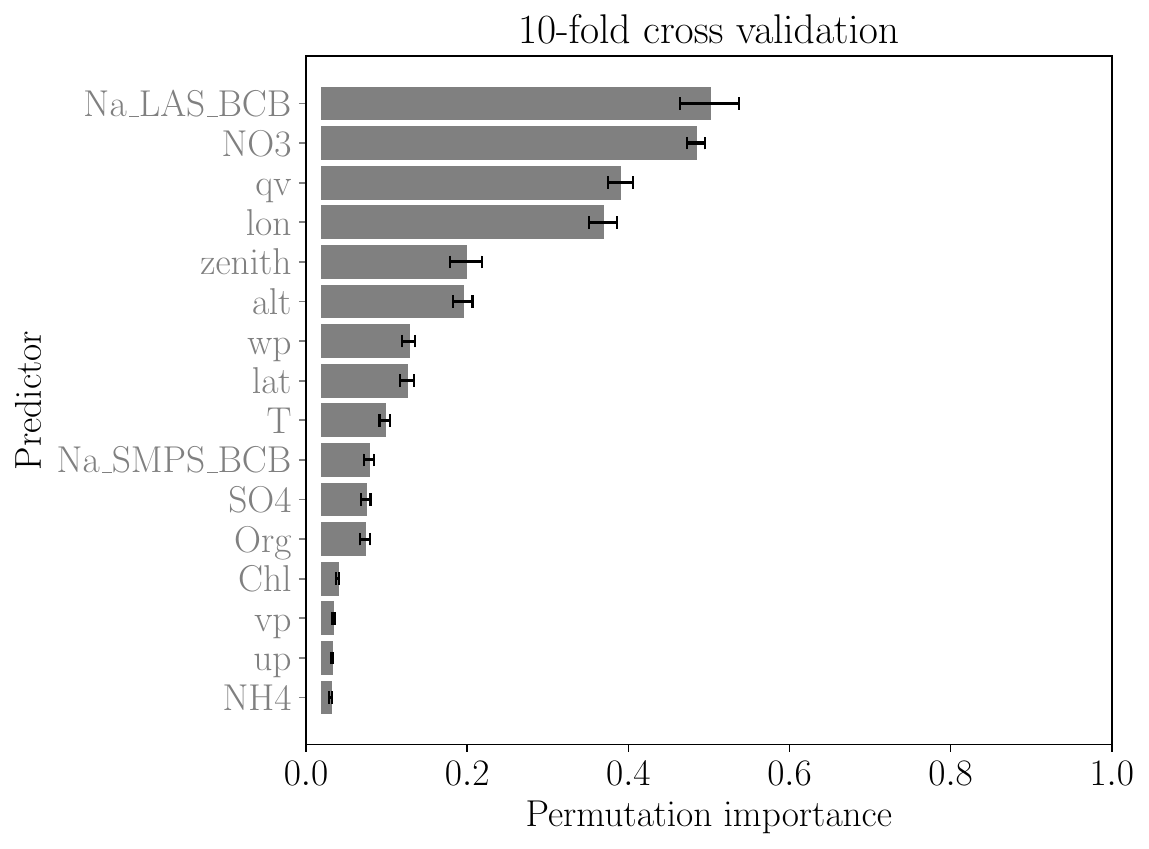}
          \label{fig:Nc_predicted_obs-b}
    \end{subfigure}
    \begin{subfigure}[b]{0.32\textwidth}
          \subcaption{}
          \includegraphics[width=\textwidth]{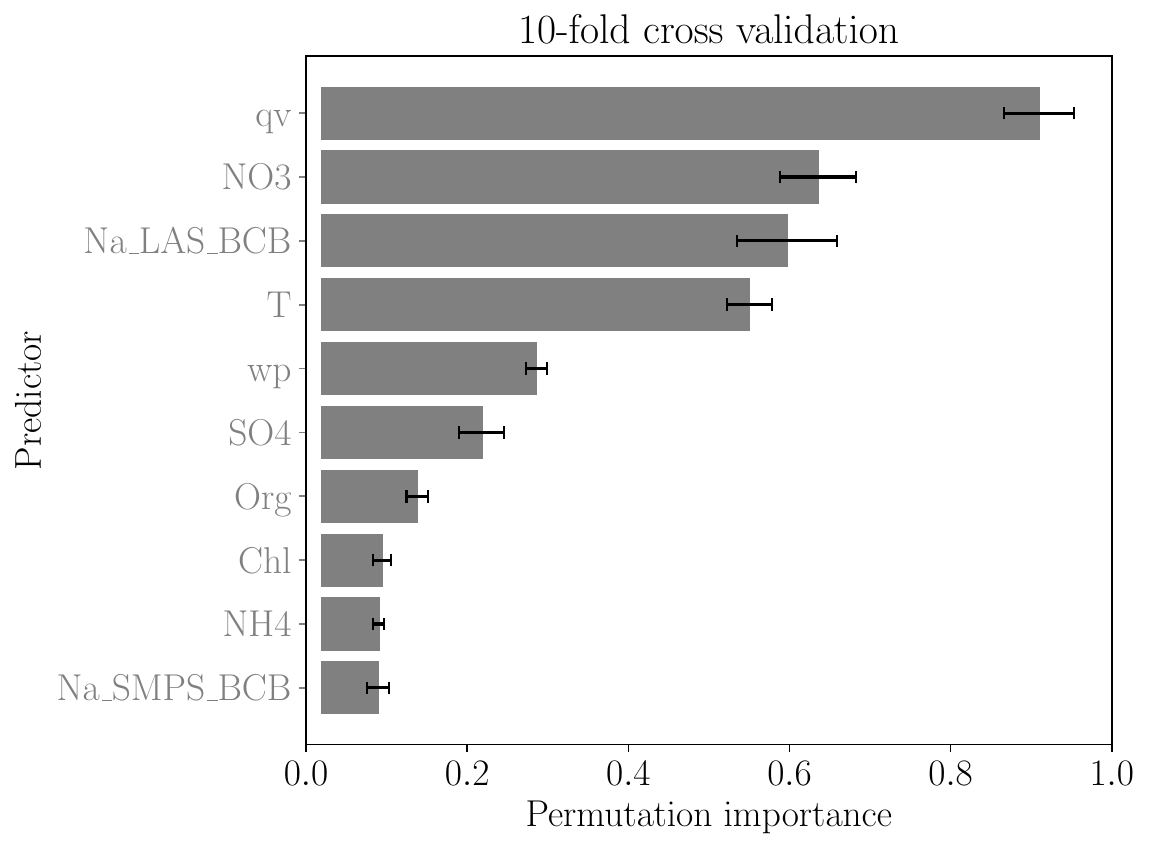}
          \label{fig:Nc_predicted_obs-e}
    \end{subfigure}
    \begin{subfigure}[b]{0.32\textwidth}
          \subcaption{}
          \includegraphics[width=\textwidth]{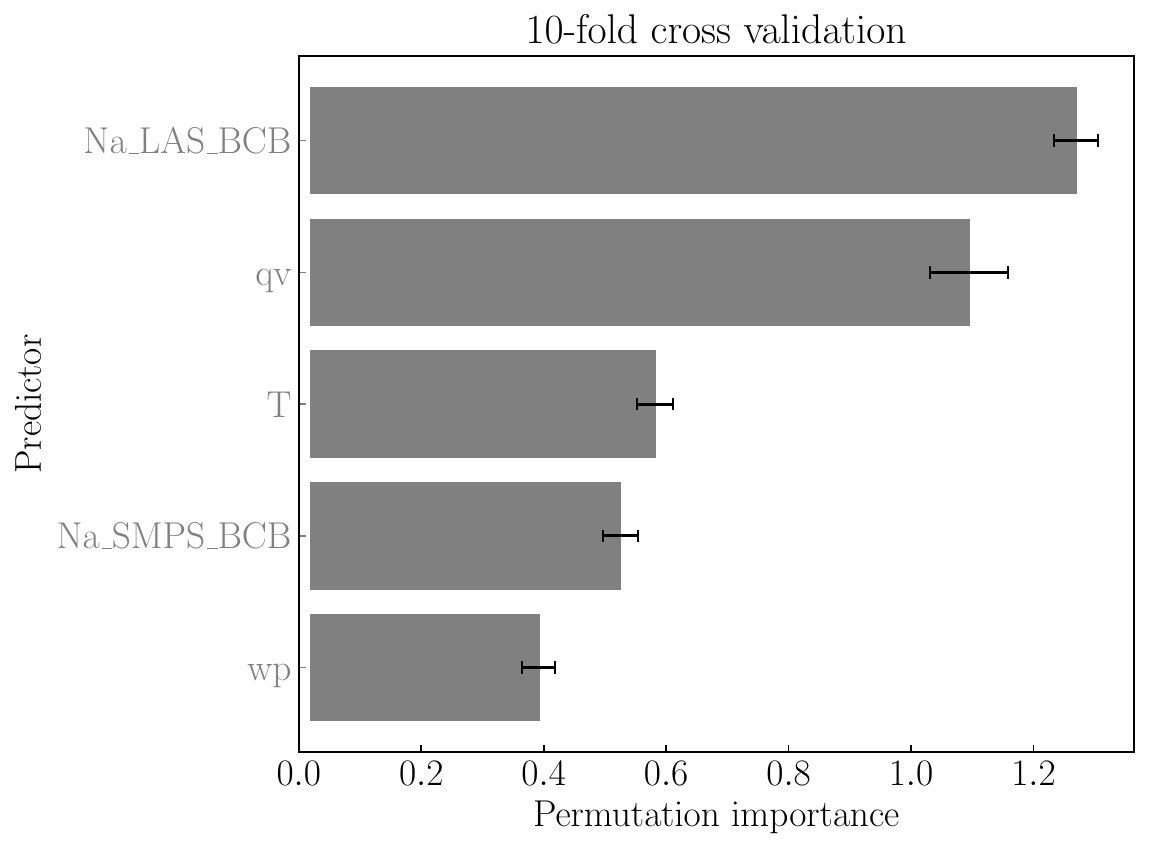}
          \label{fig:Nc_predicted_obs-h}
    \end{subfigure}
    %---MC---
    \begin{subfigure}[b]{0.32\textwidth}
          \subcaption{}
          \includegraphics[width=\textwidth]{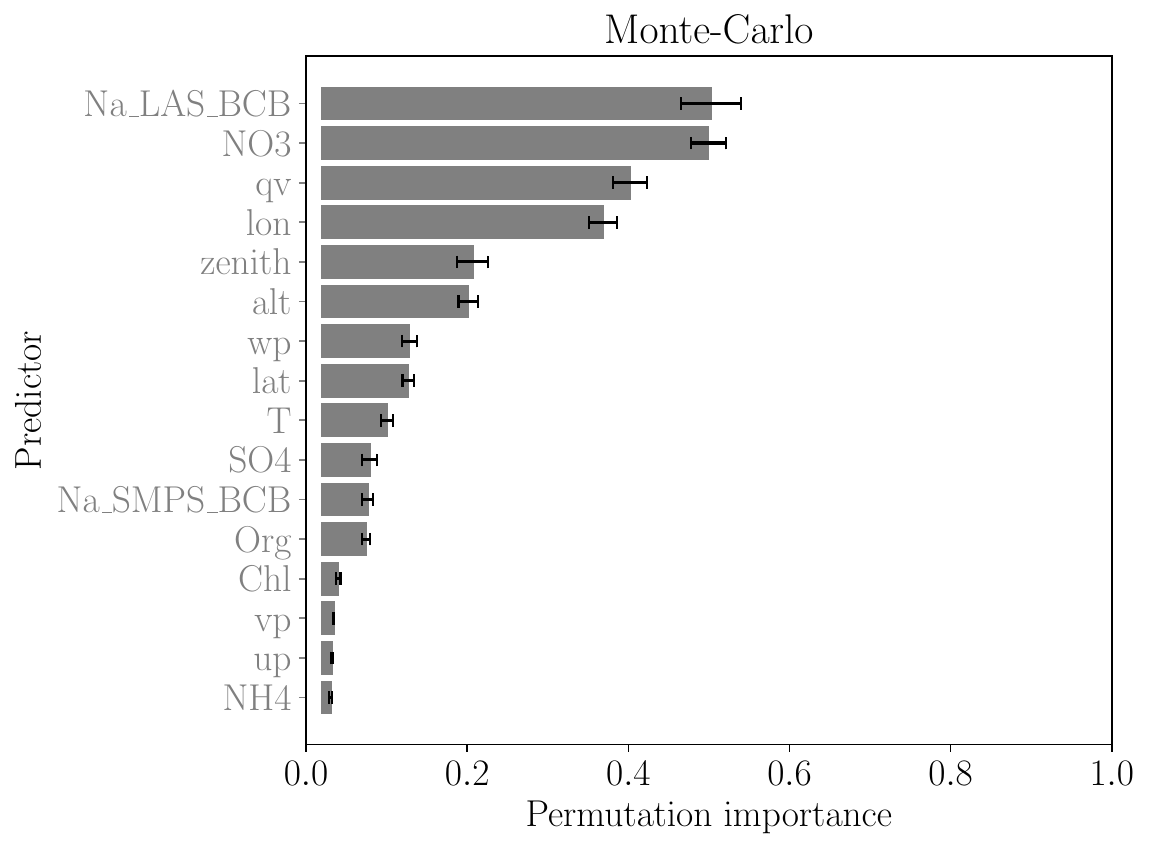}
          \label{fig:Nc_predicted_obs-c}
    \end{subfigure}
    \begin{subfigure}[b]{0.32\textwidth}
          \subcaption{}
          \includegraphics[width=\textwidth]{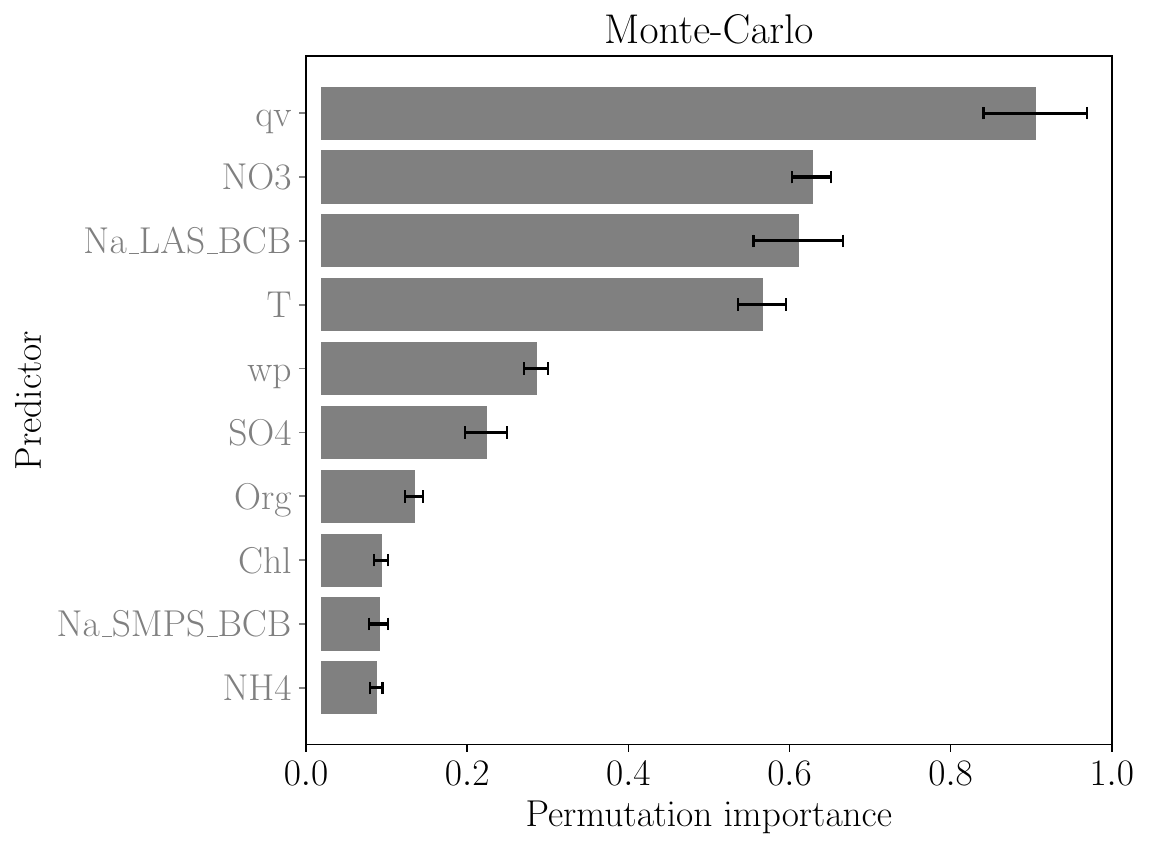}
          \label{fig:Nc_predicted_obs-f}
    \end{subfigure}
    \begin{subfigure}[b]{0.32\textwidth}
          \subcaption{}
          \includegraphics[width=\textwidth]{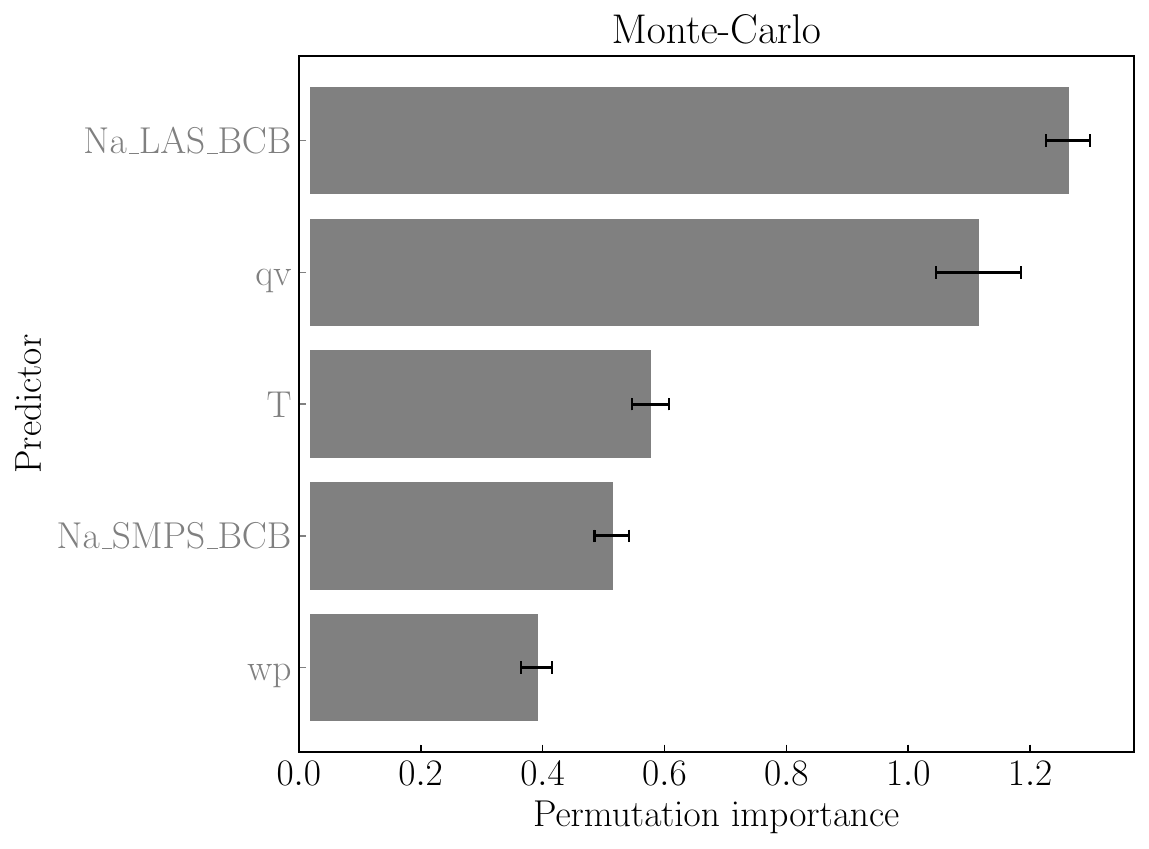}
          \label{fig:Nc_predicted_obs-i}
    \end{subfigure}
%---
\end{center}
\caption{Binned scatter-plot of \nco and \ncp from different predictors 
  (a): $N_c = \mathcal{G}(m_\mathcal{X}, N_a, w^\prime, u^\prime, v^\prime,
T, q_v, \bm{x}, \theta_z)$, i.e., all available measurements; (b) $N_c = \mathcal{G}(m_\mathcal{X}, N_a, w^\prime,
T, q_v)$, and (c) $N_c = \mathcal{G}(N_a, w^\prime,
T, q_v)$. Color bar shows the counts of data points in each hexagonal bin. The red dashed line represents the one-to-one line. The solid black line represents the linear regression relation $y=ax+b$ with $a$ and $b$ being the regression coefficients. The test size of the validation data is represented by $n$ in the legend of the scatter-plots.    
Error bars in the average PFI plots represent 
$\sigma$ deviation of PFI from the 10-fold and Monte-Carlo cross validations.  
}
\label{fig:Nc_predicted_obs}
\end{figure*}

To quantify the ACI over the WNAO region climatologically, we start with
predicting $N_c$ using all physically-related measurements from the ACTIVATE
campaign. The RF model can successfully predict $N_c$ from the predictors
$m_\mathcal{X}, N_a, w^\prime, u^\prime, v^\prime, T, q_v,
\bm{x}, \theta_z$ with $r^2 = 0.99$ using a 20-point running-average window (see detailed discussion section~\ref{sec:obs-data}) as shown in 
\Fig{fig:Nc_predicted_obs-a}.
This is remarkable considering that the $N_c$ depends on highly nonlinear and multiscale processes.
A natural question arises as to what is the relative importance of each predictor for estimating $N_c$.
As shown in
\Fig{fig:Nc_predicted_obs-b} and \Fig{fig:Nc_predicted_obs-c} , the dominant predictors are number
concentration of large-size ($\ge 100$ nm) mode aerosol particles $N_{a, \rm LAS}$, mass fraction
of nitrate $m_{\rm NO_3}$, and water vapor mixing ratio $q_v$. The importance of all three quantities to the $N_c$ prediction is
consistent with our physical understanding. Less straightforward is the importance of $m_{\rm NO_3}$, as the dominant anthropogenic 
chemical component contributing to aerosol activation is $\rm SO_4$
conventionally. In the ACTIVATE study region, sulfate and organics are the most dominant submicron species, with the latter having more of an offshore gradient as compared to sulfate which has strong influence from ocean biogenic emissions such as dimethylsulfide even over the remote ocean . However, we note that the seasonal variation of the $m_{\rm NO_3}$ follows that of $N_c$ over the WNAO region \citep{Dadashazar2022OrganicData}. Even though nitrate may not be as abundant by mass as sulfate and organics in any given season over the WNAO \citep{Dadashazar2022OrganicData}, it thermodynamically favors colder conditions, which is why it is the only species (versus sulfate and organics) exhibiting higher absolute concentrations in winter as compared to other seasons \citep{Corral2022DimethylamineOcean}, which may explain its strong association with $N_c$. 

So far, we have showed that the RFM model is able to successfully predict $N_c$ from all the available measured predictors and is able to identify top contributors to the $N_c$ prediction, which motivates us to explore whether this holds with fewer physically-motivated predictors in the spirit of dimension reduction. We first predict $N_c$ by removing the physically less important or covariant quantities, the horizontal wind speed $u \& v$, geo-coordinate $\bm{x}$, and the zenith angle $\theta_z$, from the predictors pool. This again yields a successful prediction of $N_c$ with $r^2=0.95$ (\Fig{fig:Nc_predicted_obs-d}) and retains $N_{a, \rm LAS}$, $m_{\rm NO_3}$, and $q_v$ as the main contributors (\Fig{fig:Nc_predicted_obs-e} and \Fig{fig:Nc_predicted_obs-f}). 
Ideally, the $N_c$ would be determined by chemical components of aerosols $m_\mathcal{X}$, aerosol number concentration $N_a$, the vertical component of turbulence $w^\prime$, and the thermodynamics ($T$ and $q_v$) according to the K\"ohler theory \citep{Kohler1936TheDroplets}. However, knowing the mass fraction of $\mathcal{X}$ is challenging for numerical models and observations. We therefore drop $m_\mathcal{X}$ from the predictor pool and predict $N_c$ only from $N_a$, $T$, $q_v$, and $w^\prime$. The prediction of $N_c$ and variable importance is again remarkably successful (\Fig{fig:Nc_predicted_obs-g}, \Fig{fig:Nc_predicted_obs-h}, and \Fig{fig:Nc_predicted_obs-i}).

\subsection{A stochastic nature of aerosol-cloud interactions}
\label{sec:stochastic}

\begin{figure*}[t!]\begin{center}
%---subfigure---
    \begin{subfigure}[b]{0.32\textwidth}
          \subcaption{}
          \includegraphics[width=\textwidth]{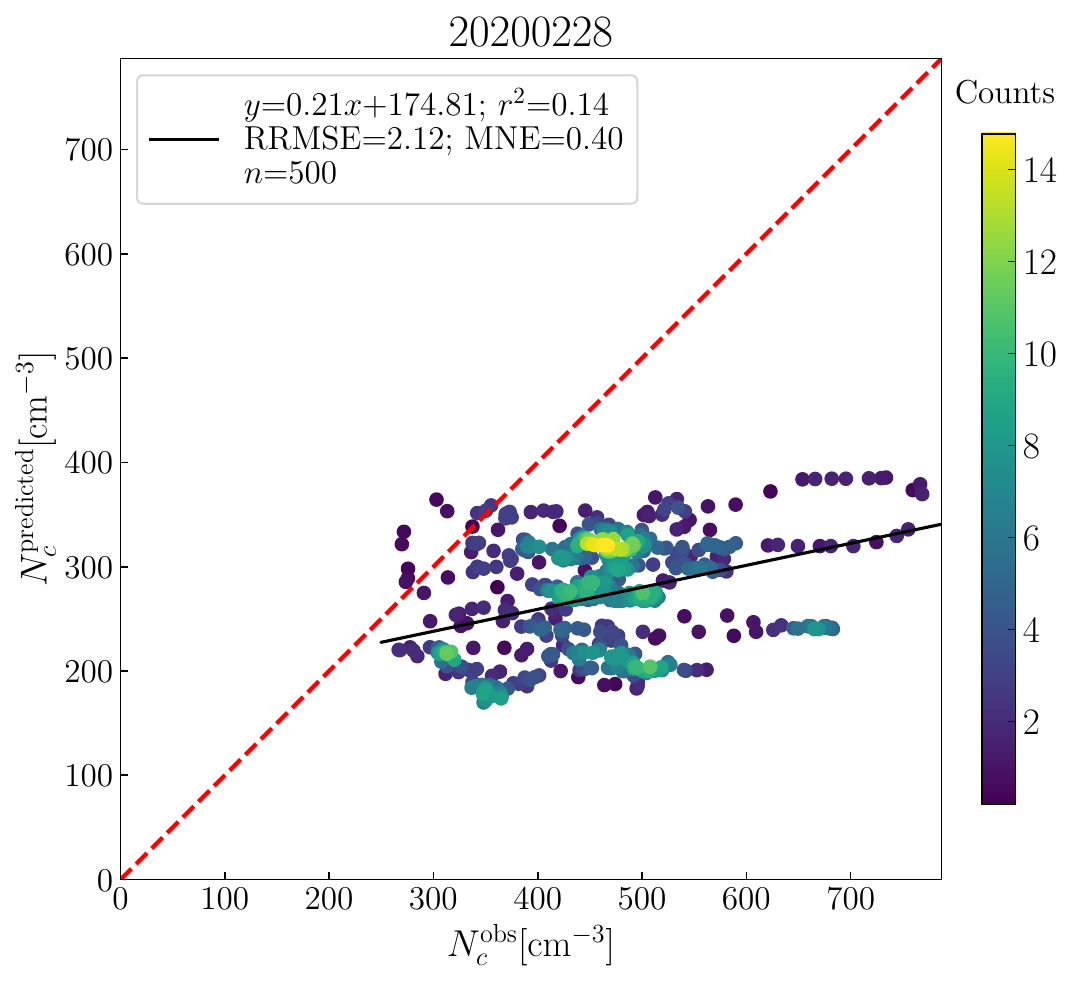}
          \label{fig:Nc_predicted_obs_20200301-a}
    \end{subfigure}
    \begin{subfigure}[b]{0.32\textwidth}
          \subcaption{}
          \includegraphics[width=\textwidth]{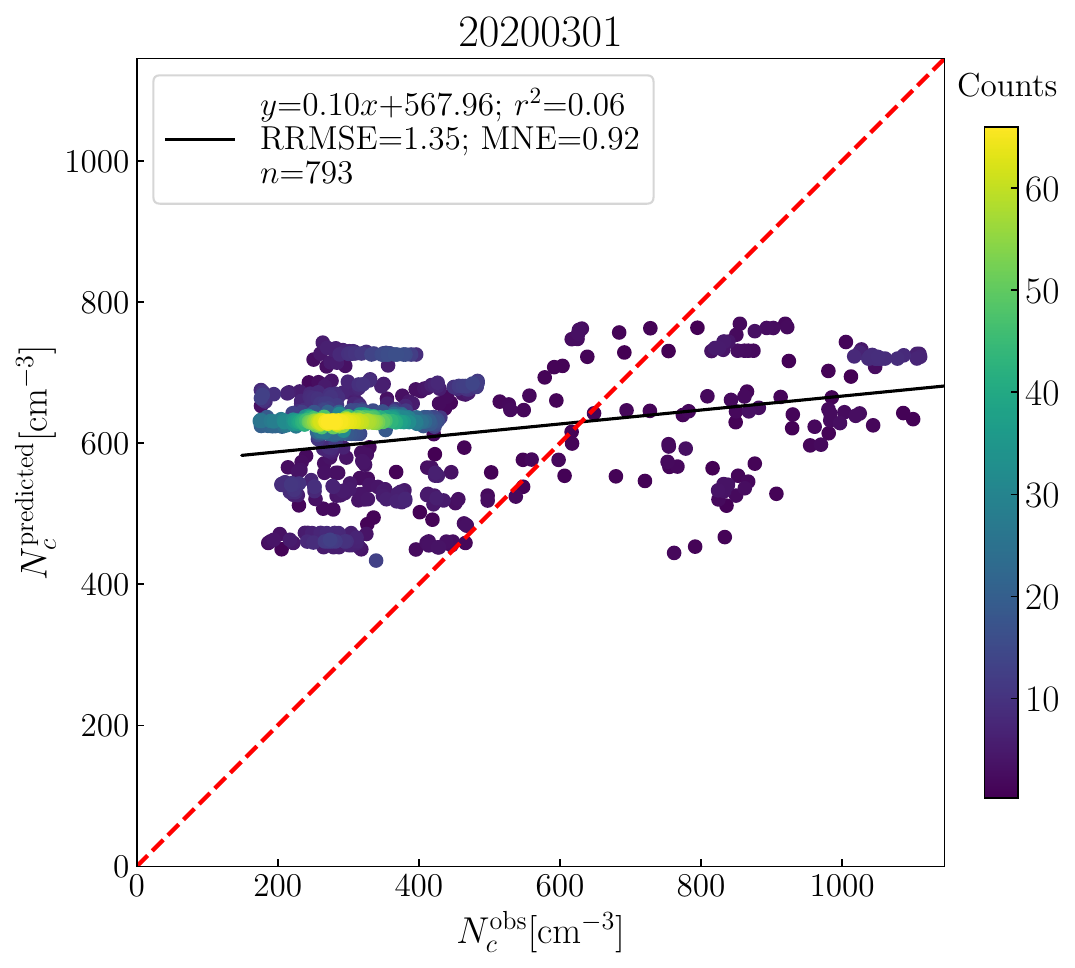}
          \label{fig:Nc_predicted_obs_20200301-b}
    \end{subfigure}
    \begin{subfigure}[b]{0.32\textwidth}
          \subcaption{}
          \includegraphics[width=\textwidth]{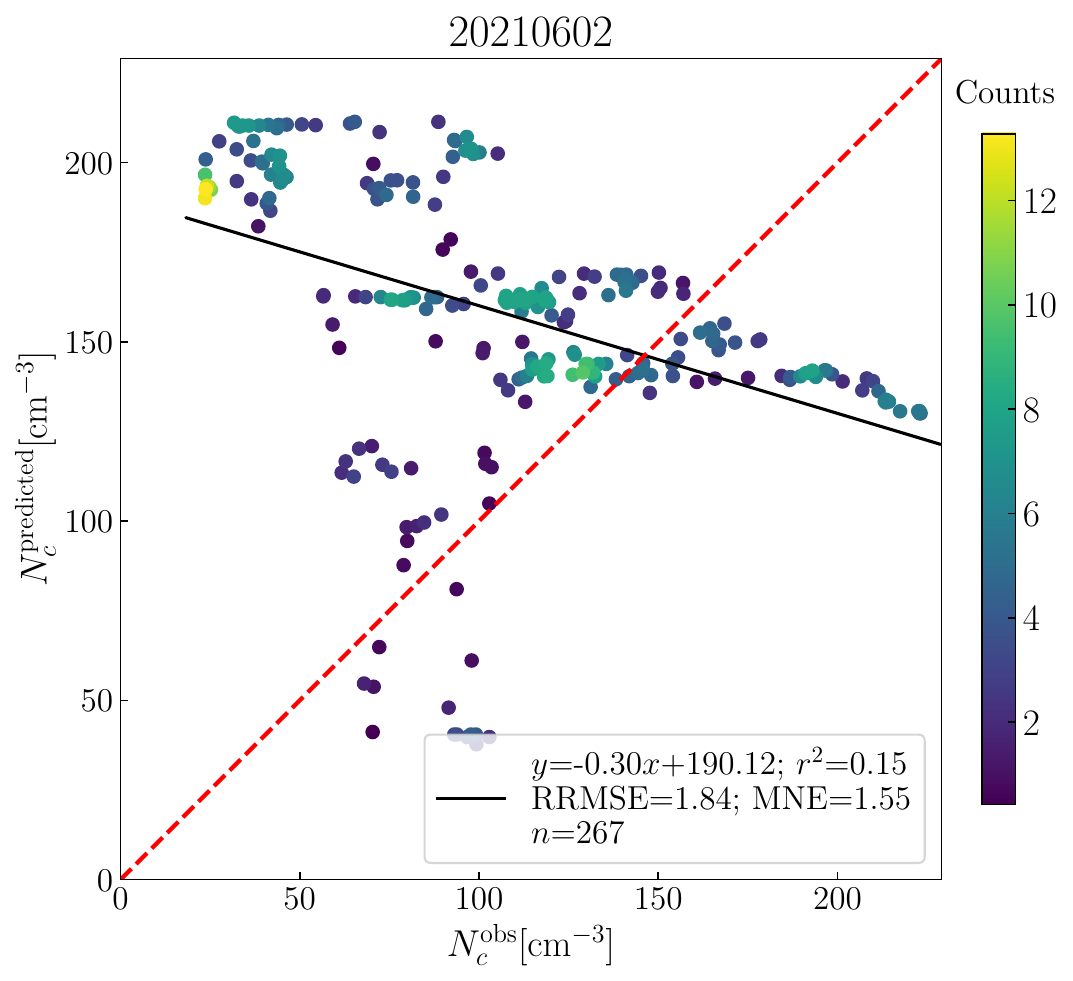}
          \label{fig:Nc_predicted_obs_20200301-c}
    \end{subfigure}
    \begin{subfigure}[b]{0.32\textwidth}
          \subcaption{}
          \includegraphics[width=\textwidth]{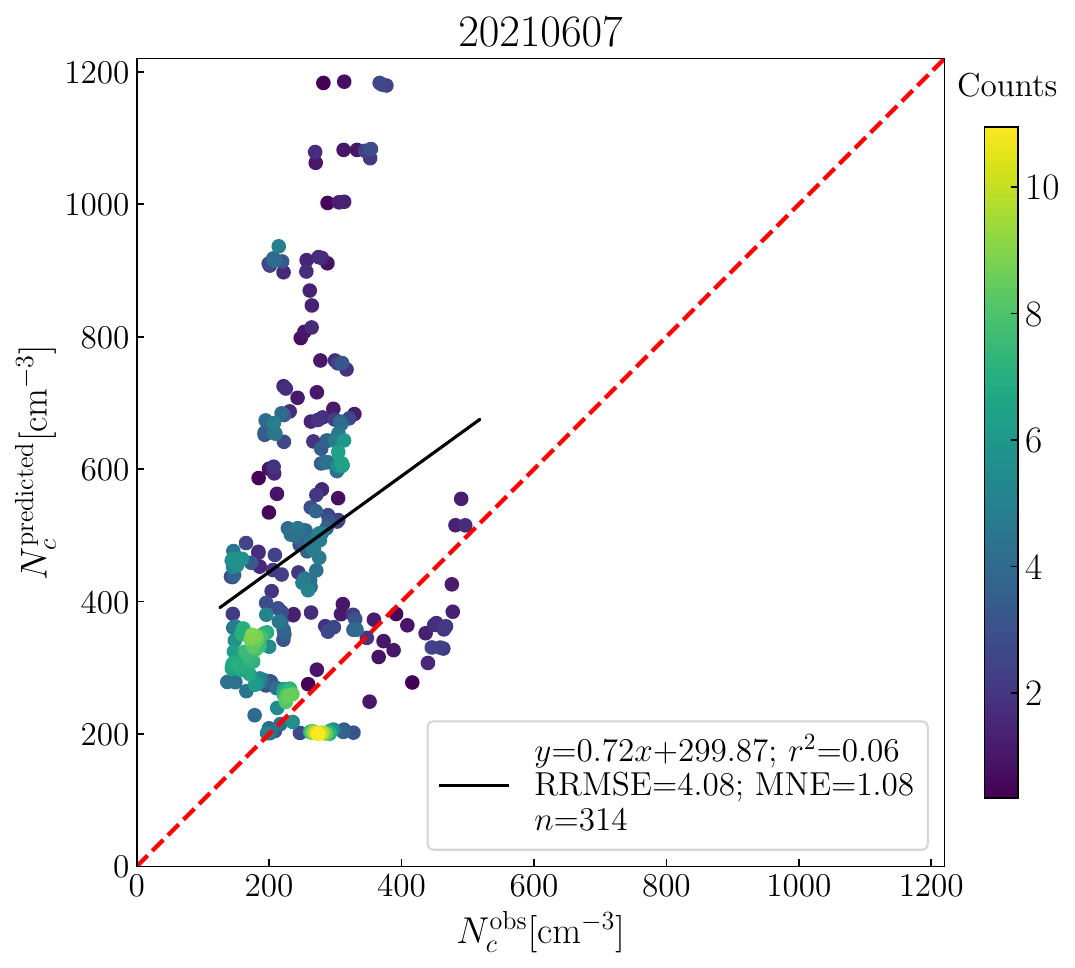}
          \label{fig:Nc_predicted_obs_20200301-d}
    \end{subfigure}
    \begin{subfigure}[b]{0.32\textwidth}
          \subcaption{}
          \includegraphics[width=\textwidth]{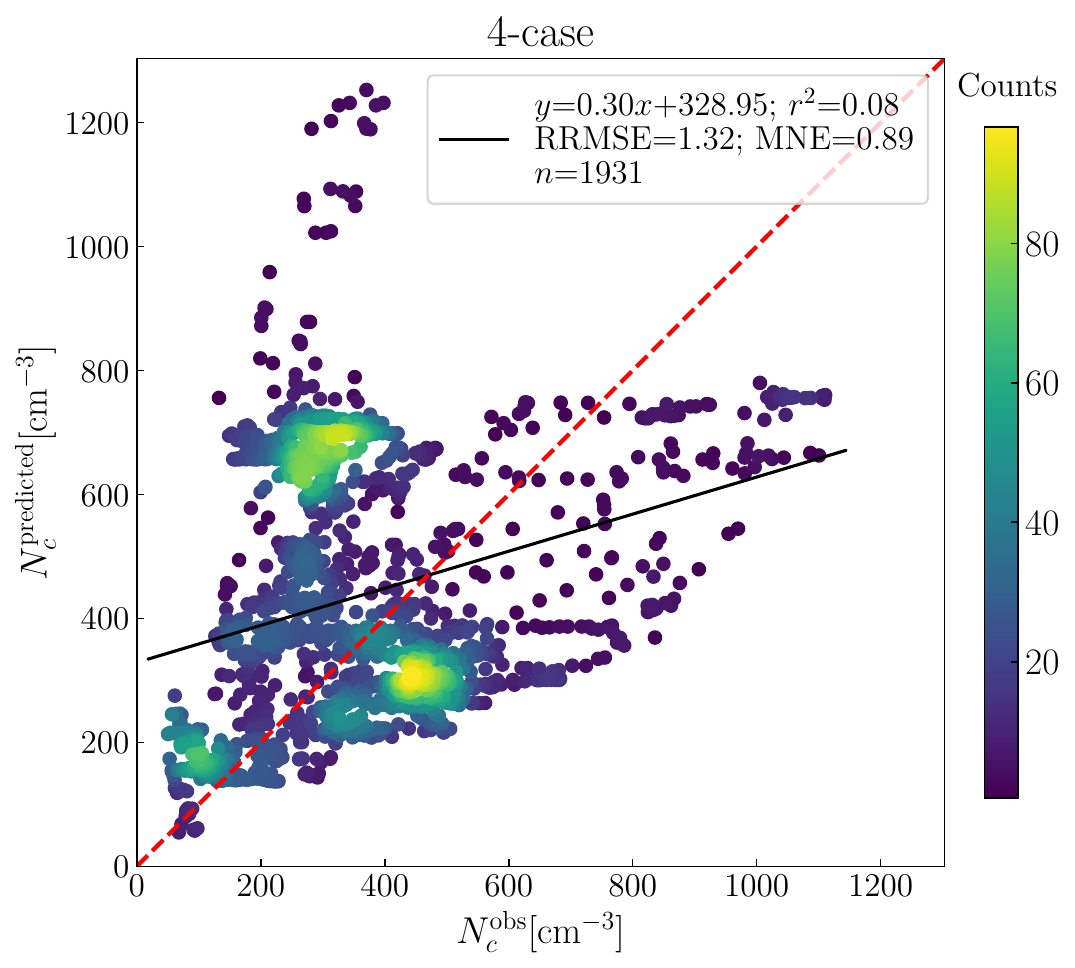}
          \label{fig:Nc_predicted_obs_20200301-e}
    \end{subfigure}
\end{center}
\caption{$N_c$ predicted from the individual observed cases (01 Mar and 28 Feb 2020 and 02/07 June 2021)
using the observation-RF (no corresponding observed case in the training dataset) is compared to the FCDP-$N_c$. The same averaging strategy is applied to the validation data for each case.}
\label{fig:Nc_predicted_obs_20200301}
\end{figure*}

\begin{figure*}[t!]\begin{center}
\begin{overpic}[width=\textwidth]{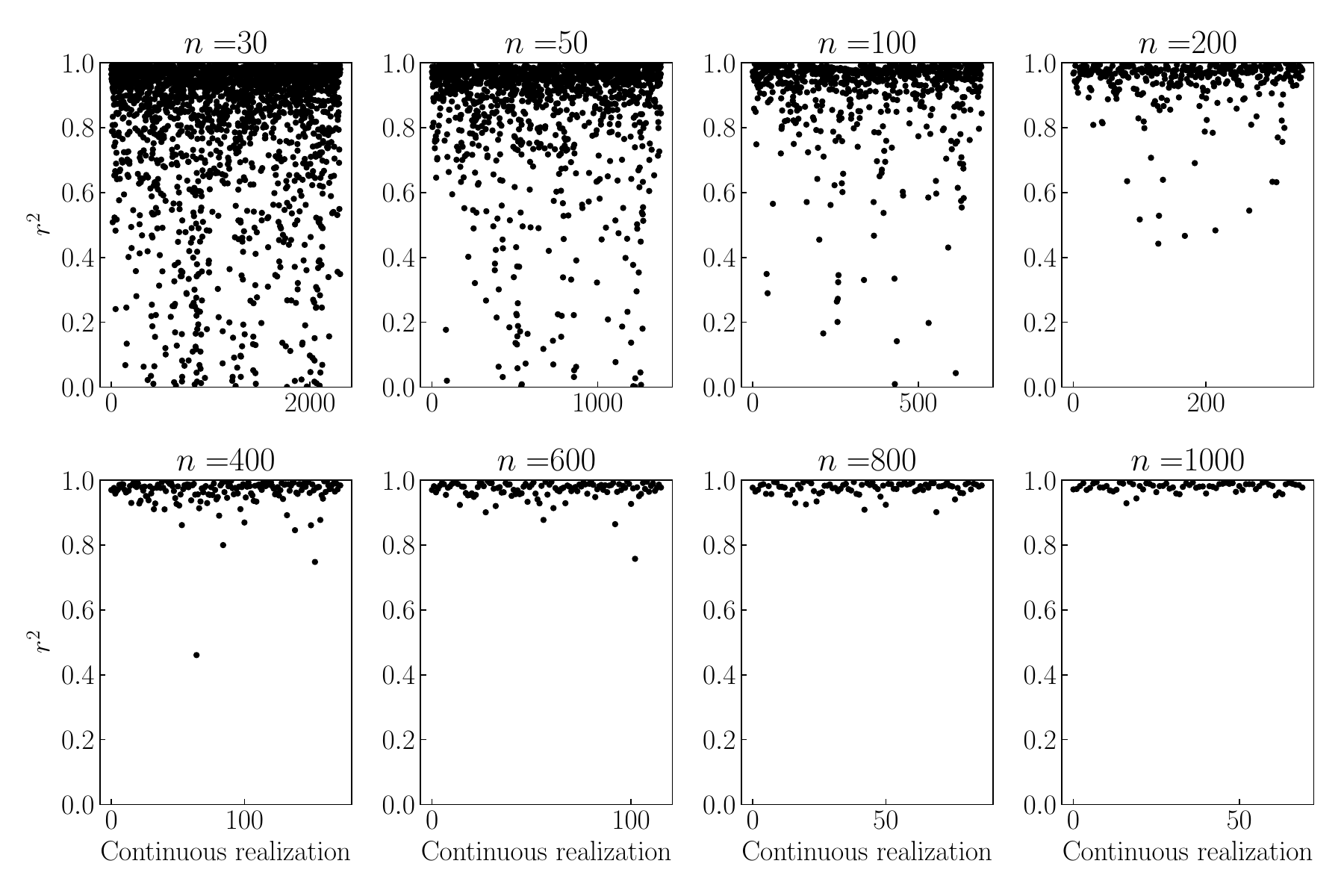}\end{overpic}
\end{center}
\caption{$r^2$ of $N_c = \mathcal{G}(N_a, w^\prime, T, q_v)$ predictions from continuous sampling with different sampling size $n$ using the full-data observation-RF. Note that the samples are drawn from running-averaged dataset with a window size of 20 data points.}
\label{fig:r2_continuous}
\end{figure*}

\begin{figure*}[t!]\begin{center}
\begin{overpic}[width=0.32\textwidth]{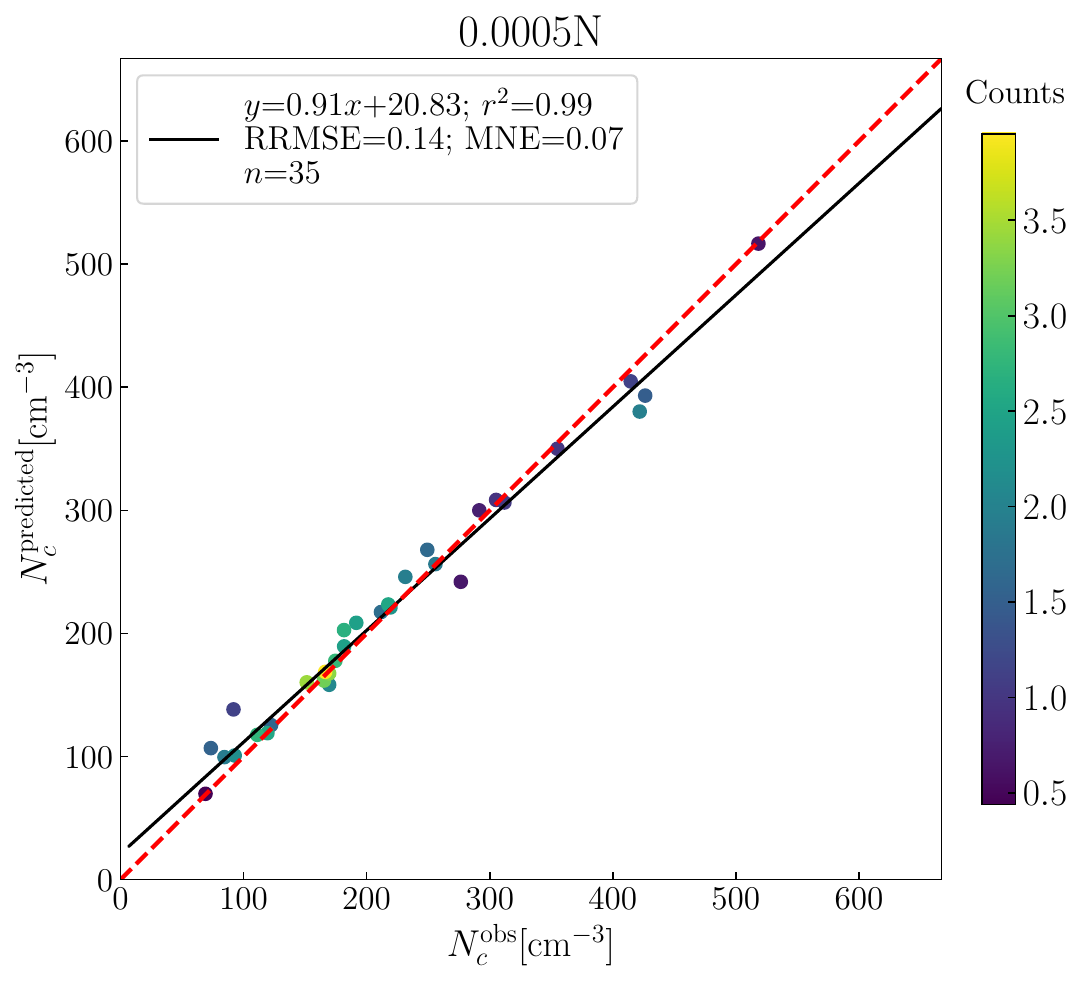}\end{overpic}
\begin{overpic}[width=0.32\textwidth]{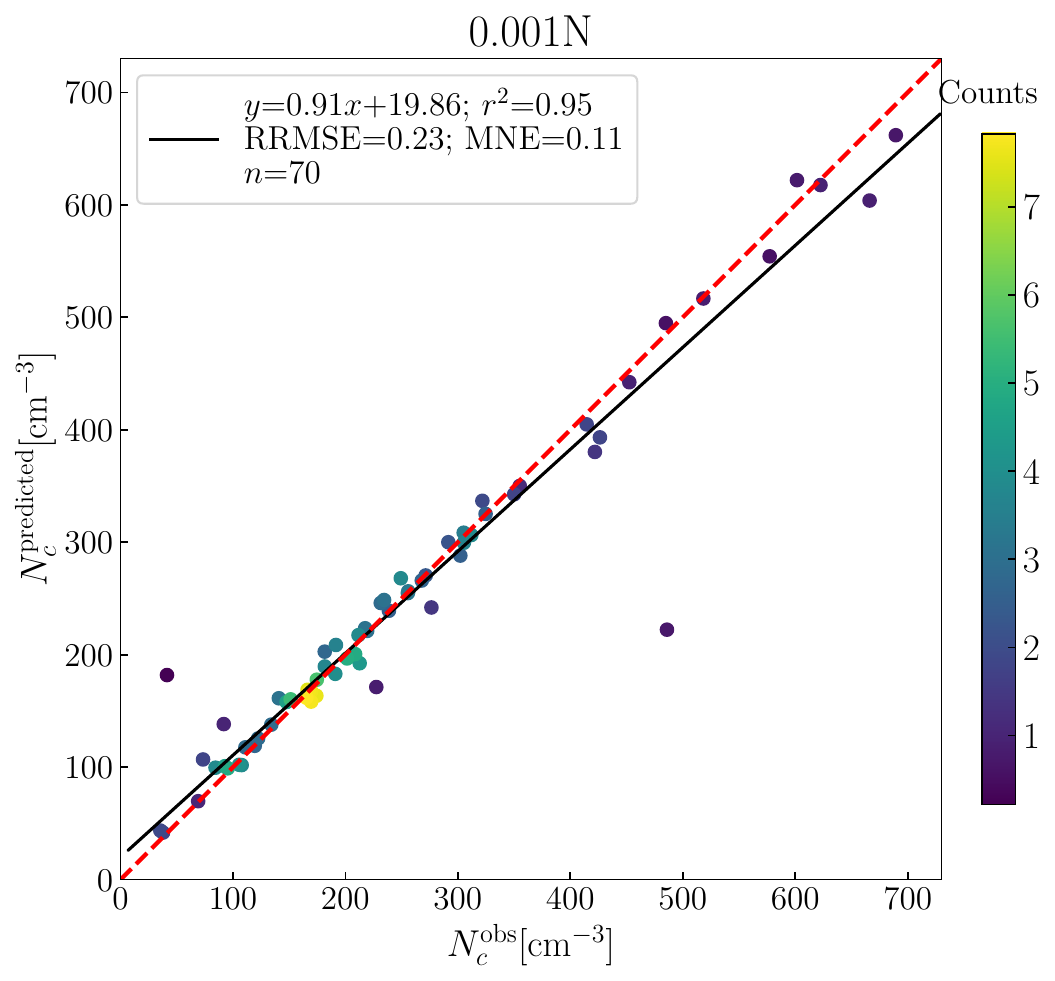}\end{overpic}
\begin{overpic}[width=0.32\textwidth]{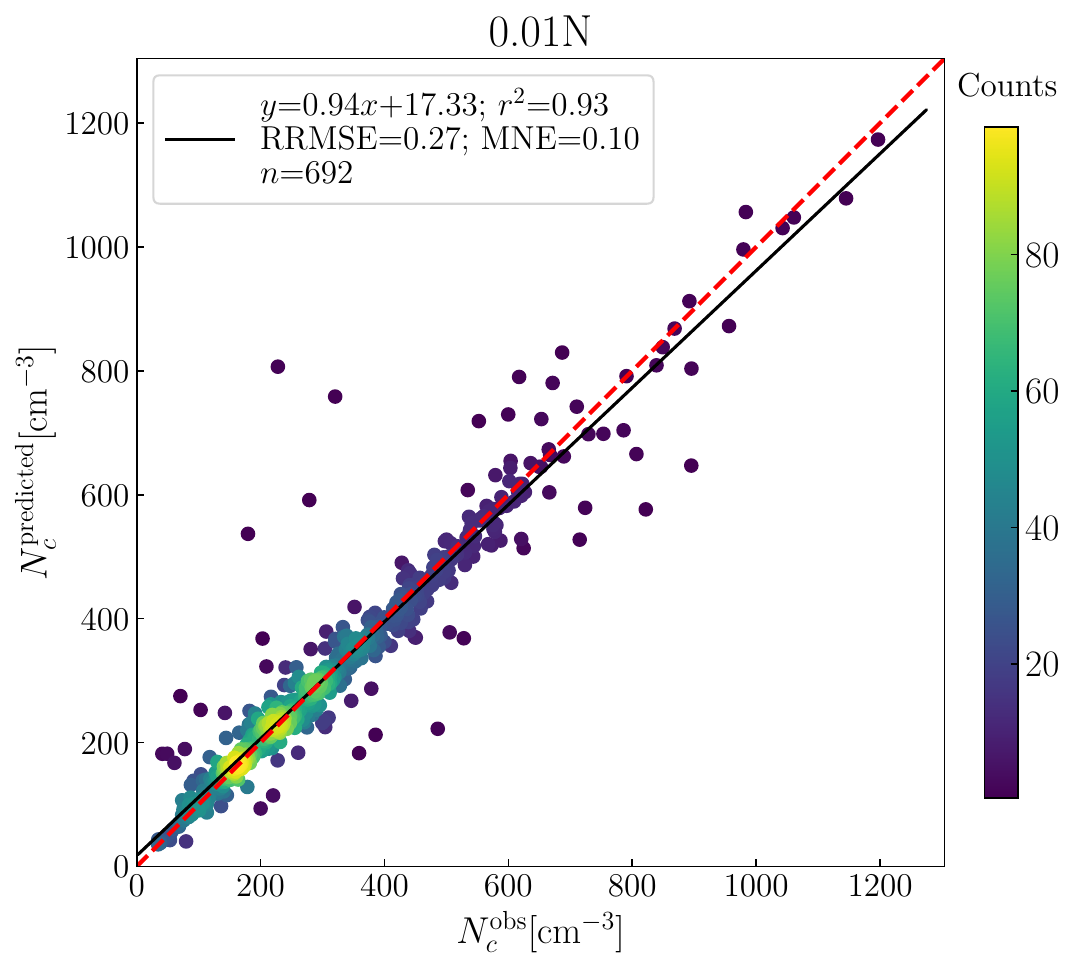}\end{overpic}
\begin{overpic}[width=0.32\textwidth]{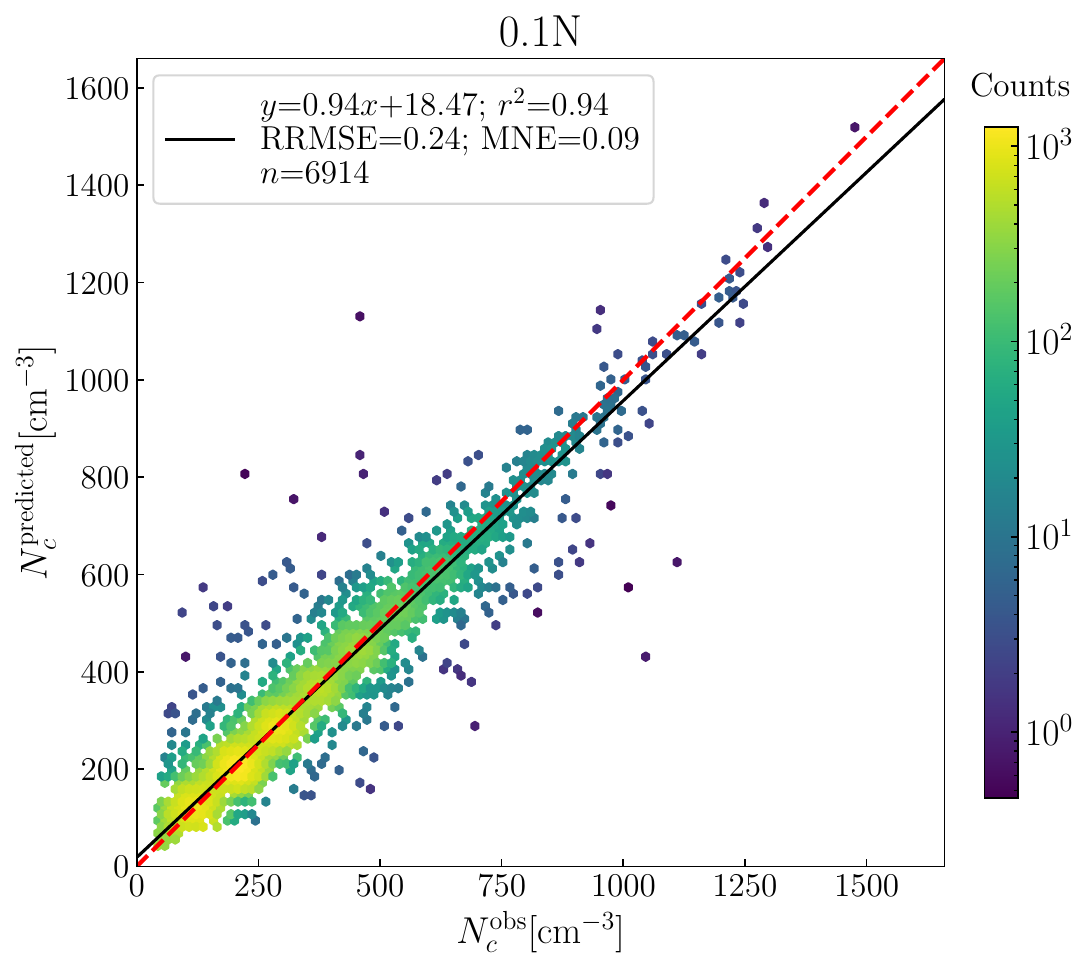}\end{overpic}
\begin{overpic}[width=0.32\textwidth]{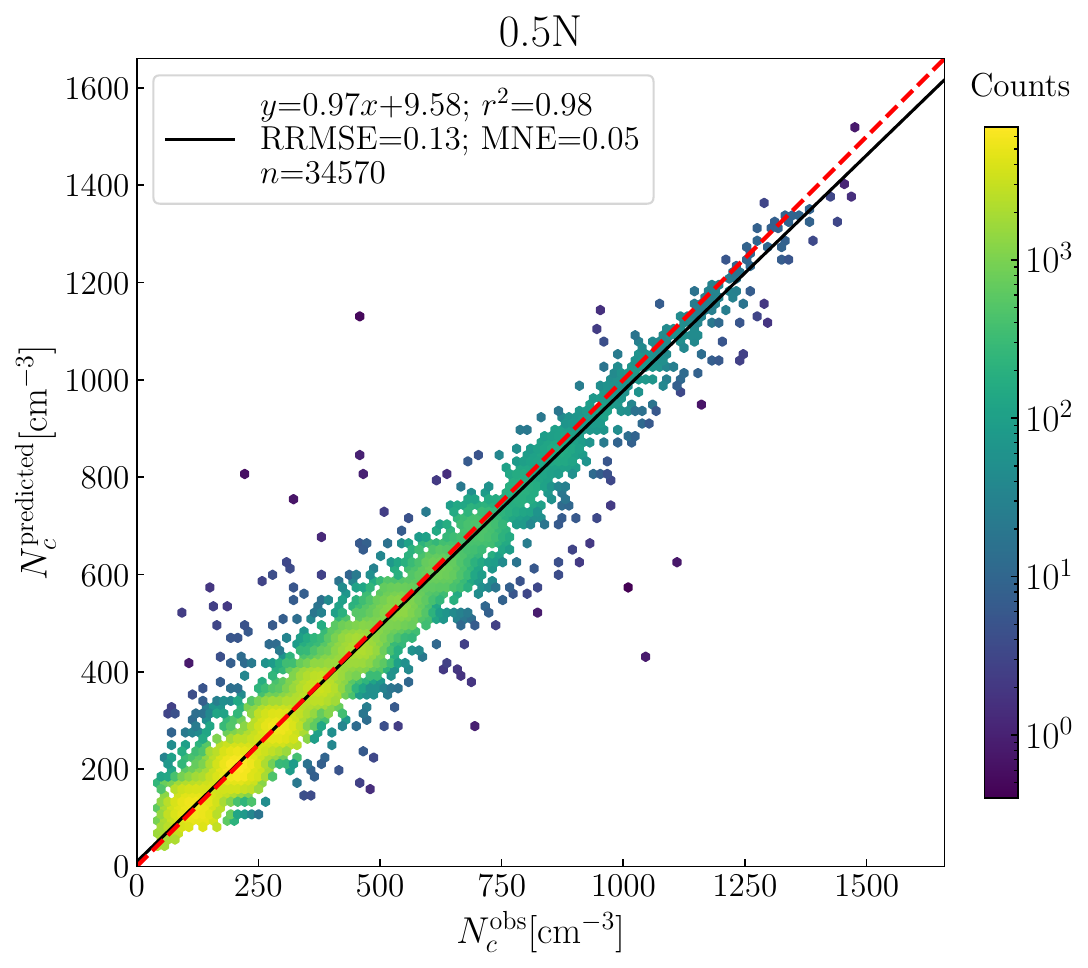}\end{overpic}
\begin{overpic}[width=0.32\textwidth]{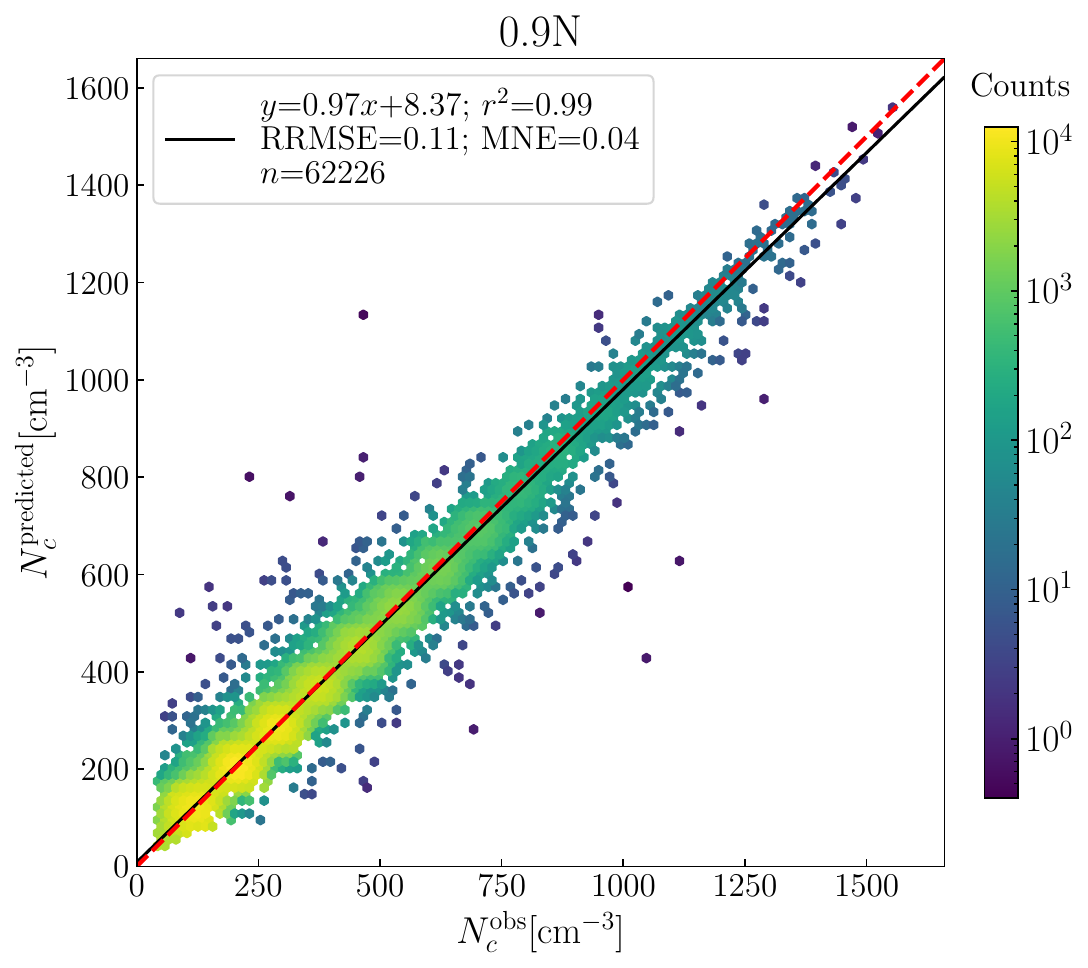}\end{overpic}
\end{center}
\caption{Test of the input sample size for the same observation-RF with the
full dataset. The same features $N_a, w^\prime, T, \text{and}\, q_v$ are used as in
\Fig{fig:Nc_predicted_obs_20200301} and \Fig{fig:r2_continuous}. The data points are randomly selected from
the averaged full dataset.}
\label{fig:Nc_predicted_obs_size}
\end{figure*}

With the successful prediction of $N_c$ from the observational data in hand, we can
now examine whether and how the $N_c$ prediction is stochastic or deterministic and its implications
on our understanding of multiscale ACI processes. Recall that the observation-RFM is trained and
validated using 3-year in-situ measurements, which represents a statistical
$N_c$ prediction for the WNAO domain over multiple years. The stochastic or deterministic nature of ACI metrics can be
pursued by examining the predictive ability of the observation-RFM for individual flights. This is achieved by predicting $N_c$ for single-day
events using observation-RFM trained and validated from the 3-year in-situ
measurements excluding the targeted flights. Such observation-RFM fails to predict 
$N_c$ for all four specific events we choose to examine, including two wintertime cold-air outbreak cases observed on 01 Mar
and 28 Feb 2020, respectively, and two summertime cumulus cases on 02 and 07 June 2021, respectively, as shown in
\Fig{fig:Nc_predicted_obs_20200301-a}-\Fig{fig:Nc_predicted_obs_20200301-d}. We further predict $N_c$ for the
combined 4 cases using the observation-RFM trained and validated excluding
these 4 cases to consider the seasonal variations in aerosol and clouds to some extent. The observation-RFM
again fails to reproduce the observed $N_c$ from the observational inputs (\Fig{fig:Nc_predicted_obs_20200301-e}).   
To study whether this failure is case-dependent or generic, we predict $N_c$
for the continuous dataset with different sample sizes. Datasets with different
sample size $n$ represent either a single event or a few randomly selected
events. For example, $n=30$ represents a single event while $n \ge 200$
represents multiple events even though there is no fixed threshold because
individual flight events have different numbers of data points (20 s each). The
$N_c$ predictability is quite low for the $n=30$ datasets as the $r^2$ values
vary randomly from $\sim 0.0$ to $\sim 1.0$ (\Fig{fig:r2_continuous}), which
echoes the low-performance $N_c$ prediction for the specific cases in
\Fig{fig:Nc_predicted_obs_20200301}. The $r^2$ improves significantly for $n
\ge 200$, i.e., continuous multiple events (i.e., consecutive data points) for
a longer duration. This evolution of $r^2$ as a function of $n$ shows that the
$N_c$ prediction is stochastic for short timescales and is only physically and
statistically meaningful for long timescales. Namely, the cloud droplet
response to aerosols, as one of the ACI metrics, appears to be more stochastic
at the shorter timescales (or within a limited sampling area). We note that it is the spatiotemporal range of predictors that determines the $N_c$ prediction instead of the number of events. To demonstrate this, in contrary to the continuous sampling for a sub-dataset with a fixed sample size, we randomly sample a sub-dataset from the 3-year observational dataset.
The predicted $N_c$ from these randomly sampled sub-datasets reproduce the observed $N_c$ remarkably well with $r^2>0.9$ (\Fig{fig:Nc_predicted_obs_size}). $r^2 = 0.99$ even for the sub-dataset with a sample size of $n = 0.0005N \approx 35$ with $N$ being the total number of 3-year data points. The score for sub-dataset with this sample size is statistically significant as can be seen from $r^2$ for 1000 realizations (\Fig{fig:r2_size}).    

\subsection{Observation-RFM as an emulator for the LES microphysics}

\begin{figure}[t!]\begin{center}
    \begin{subfigure}[b]{0.32\textwidth}
          \subcaption{}
          \includegraphics[width=\textwidth]{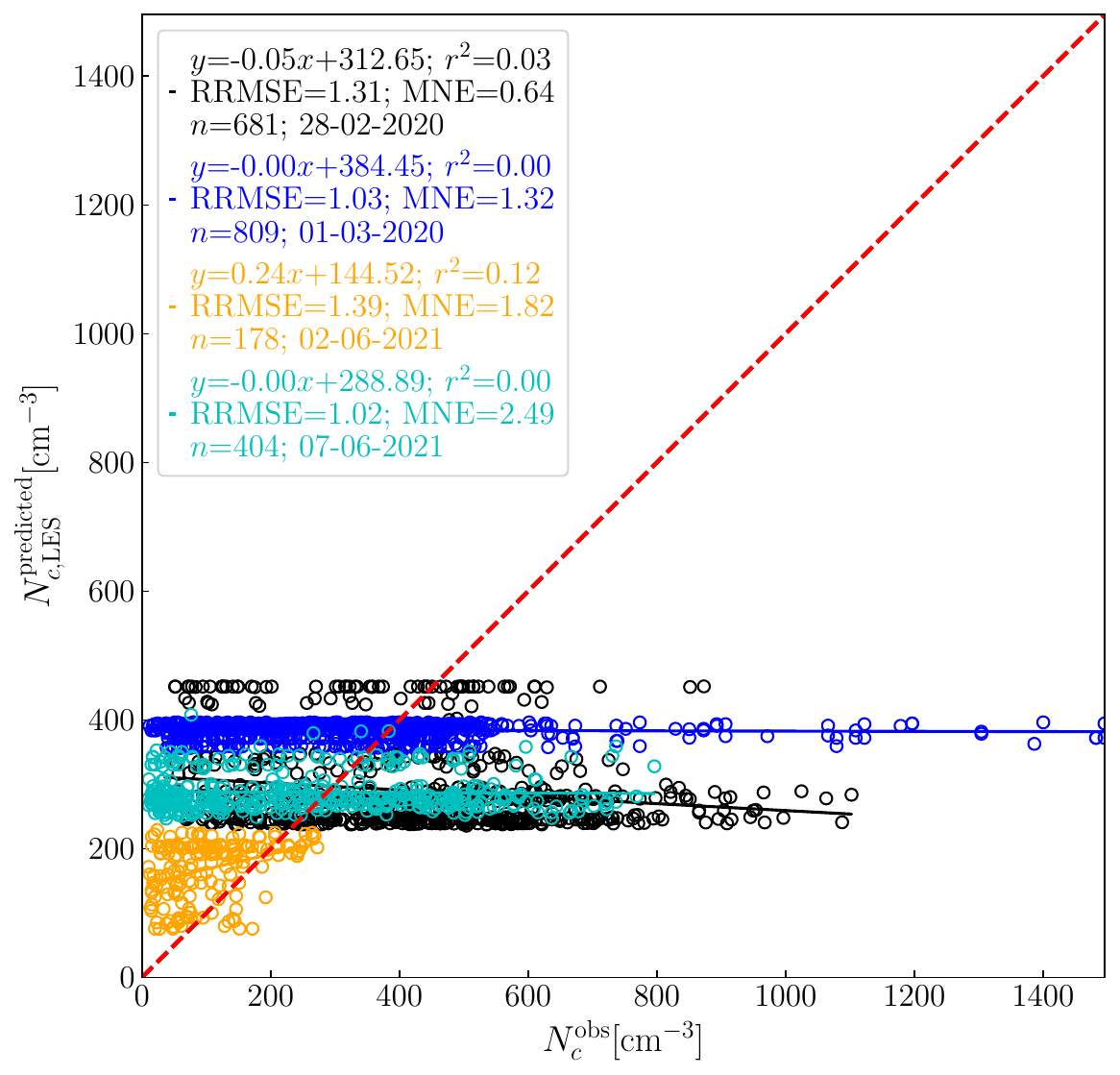}
          \label{fig:Nc21_0602-a}
    \end{subfigure}
    \begin{subfigure}[b]{0.32\textwidth}
          \subcaption{}
          \includegraphics[width=\textwidth]{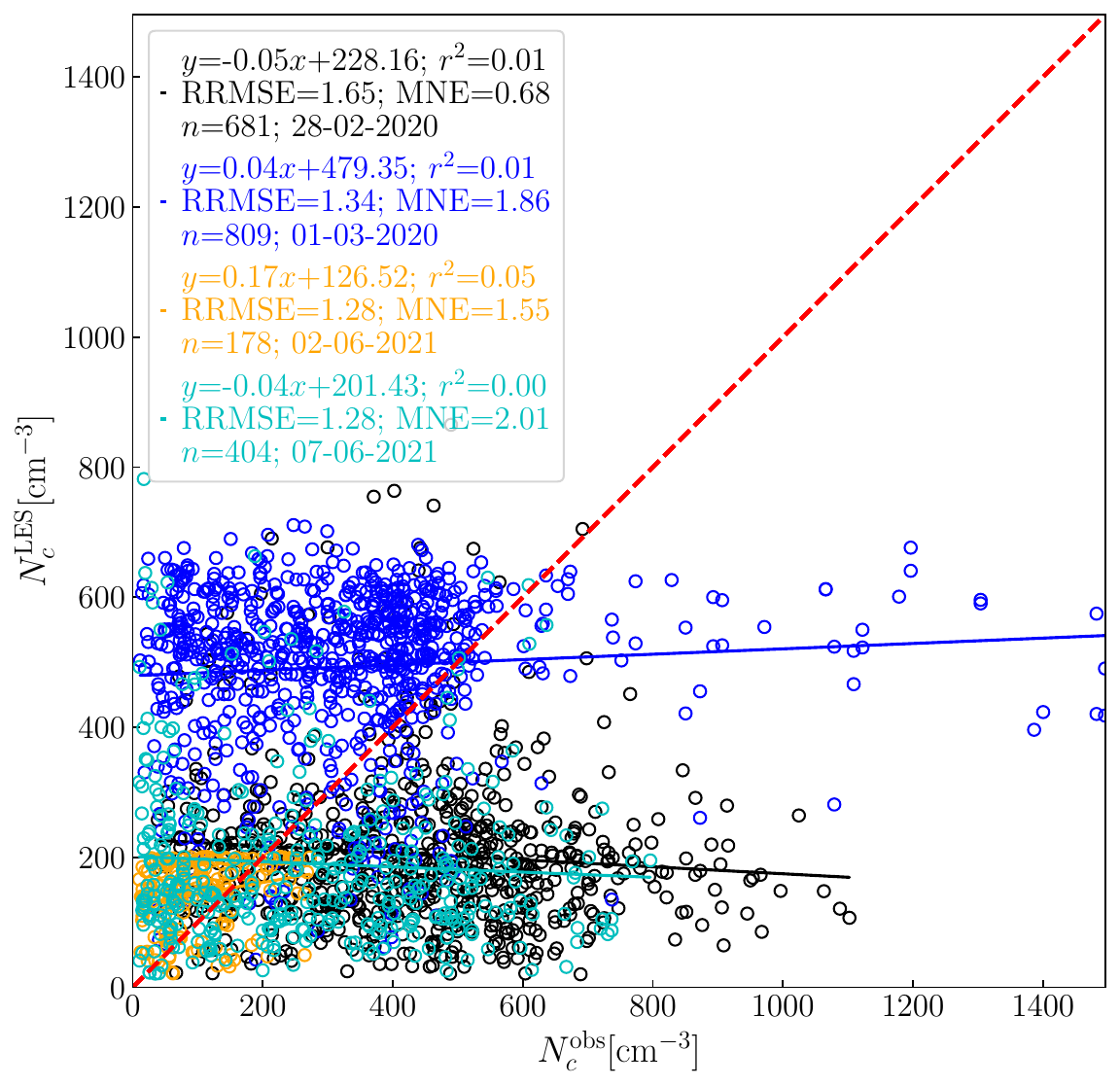}
          \label{fig:Nc21_0602-b}
    \end{subfigure}
    \begin{subfigure}[b]{0.32\textwidth}
          \subcaption{}
          \includegraphics[width=\textwidth]{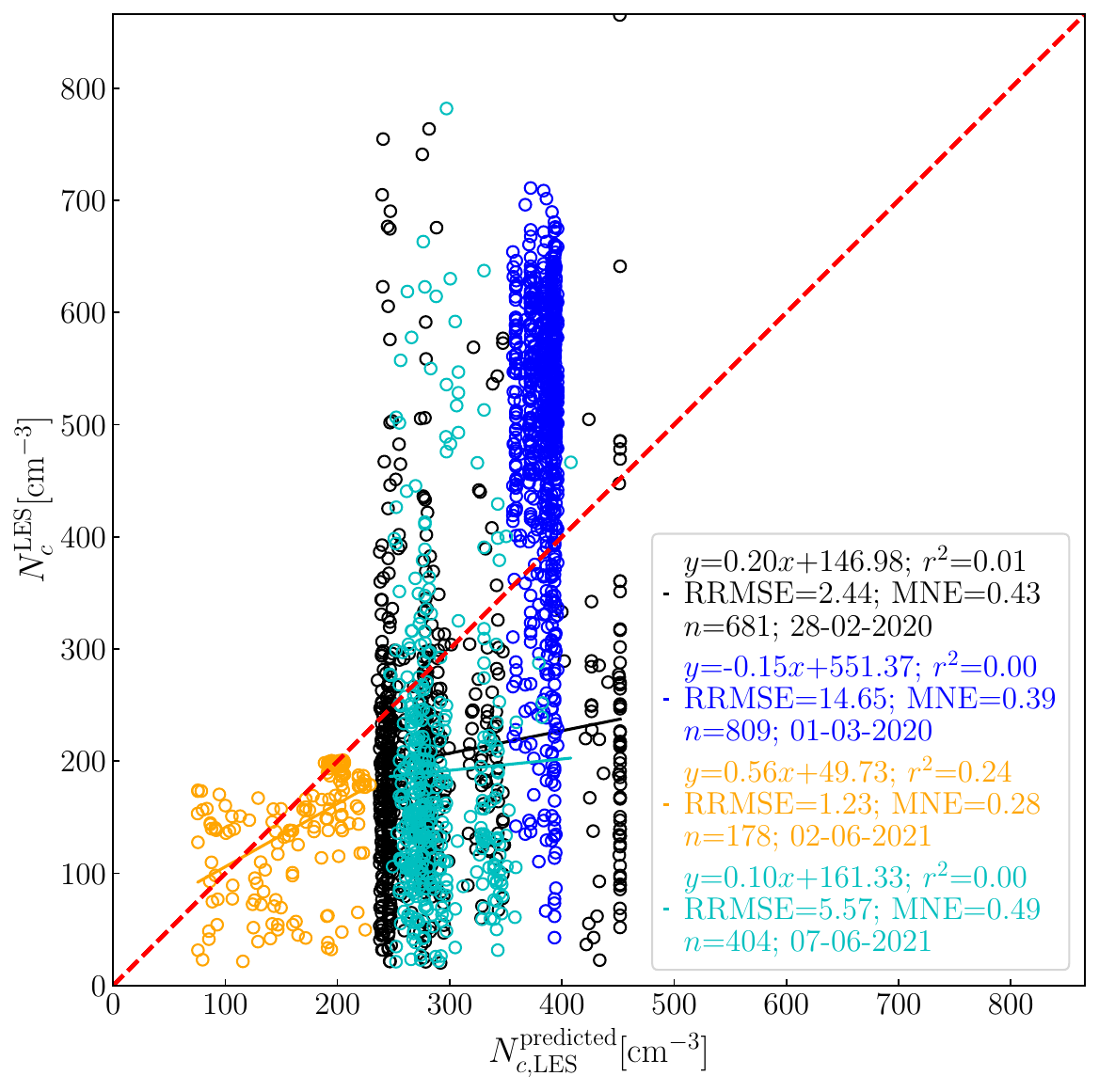}
          \label{fig:Nc21_0602-c}
    \end{subfigure}
\end{center}
\caption{$N_c$ predicted ($N_{c, \rm LES}^{\rm predicted}$) from the WRF-LES as input (LES $N_a, w^\prime, T, q_v$)
using the full-data observation-RF compared to the LES-$N_c$ ($N_c^{\rm LES}$) or FCDP-$N_c$ ($N_c^{\rm obs}$).}
\label{fig:Nc21_0602}
\end{figure}

Cloud microphysical processes for ACI are challenging to represent even in LES \citep{Li2022Large-EddyForcings, Li2023Large-EddyInteraction}. In this section, we tackle this challenge by using the observation-RFM as an emulator for the LES microphysics. Important turbulence scales for cloud formation can be resolved in the LES. In addition, we have shown that the observation-RFM can successfully predict observed $N_c$ in section~\ref{sec:obs-RFM}. It is natural to ask whether the observation-RFM as an emulator can mitigate the uncertainties of cloud-microphysics representation in LES. We use LES from two cold-air outbreak cases \citep{Li2022Large-EddyForcings, Li2023Large-EddyInteraction} and two summertime marine cumuli cases \citep{Li2023ProcessAtlantic} to represent different aerosol conditions, meteorological states, and cloud regimes over the WNAO region. $w^\prime$, $T$, $q_v$, and $N_a$ from LES are taken as predictors to predict $N_c$ using the observation-RFM. The observation-RFM fails to predict $N_c$ for the four LES cases (\Fig{fig:Nc21_0602-a}), which is consistent with the stochastic nature of ACI, as discussed in section~\ref{sec:stochastic}. 
We further evaluate the LES-$N_c$ against \nco for completeness. As shown in \Fig{fig:Nc21_0602-b}, the LES fails to reproduce the observed $N_c$ at the same flight-leg levels, as indicated in the four LES cases with $r^2 \approx 0$. This can be attributed to the following three main reasons: 1. LES with the periodic boundary conditions in horizontal directions cannot simulate the observed lateral variation of $N_c$, which makes the comparison of $N_c$ (spatiotemporally point-to-point comparison) between LES and observations challenging; 2. prescribed aerosols in the LES cannot represent the fine-scale spatiotemporal variability of aerosol size distribution; and 3. the single-value bulk hygroscopicity for all size modes is not representative. The resulting $N_c$ is expected to deviate from observations. Despite the poor performance of LES microphysics and the observation-RFM in reproducing the observed $N_c$ for individual cases, it is  still informative to compare $N_c^{\rm LES}$ and \ncp directly. They are nearly uncorrelated to each other (\Fig{fig:Nc21_0602-c}).     

\section{Discussion and conclusion}

Quantifying aerosol-cloud interactions (ACI) is very challenging due to the nonlinear multiscale nature and the incomplete understanding of related physical processes. Coarse-resolution Earth System Models only model a mean state of ACI at the model grid-scale that lacks accurate representation of physical processes and justification \citep{Morrison2020ConfrontingMicrophysics}, while the small-scale LES simulates incomplete physical processes of ACI and only offers partial representation of ACI \citep{Li2023Large-EddyInteraction, Li2023ProcessAtlantic}. An imminent question would be whether the ACI is stochastic or deterministic. Many studies \citep{Seinfeld2016ImprovingSystem, Bellouin2020BoundingChange} have alluded on this question but fallen short in explicitly formulating and answering it. In this study, we explore the stochastic nature of ACI by applying a widely used machine leaning (ML) technique, Random Forest Model (RFM), to unprecedented three-year in-situ aircraft measurements to predict $N_c$ over the western North Atlantic. Here, we focus on the response of $N_c$ to aerosols properties, thermodynamics, and turbulence. The RFM can successfully predict the climatological $N_c$ using the measured aerosol number concentration $N_a$, $w^\prime$, temperature $T$, and water vapor mixing ratio $q_v$ despite the strongly nonlinear aerosol and cloud microphysical processes, e.g., aerosol activation and condensation and collision-coalescence of cloud droplets. However, the observation-trained RFM (observation-RFM) fails to predict $N_c$ at shorter timescales that only cover a limited number of flights. This suggests that within this data-driven framework, the ACI is more stochastic at the shorter timescales. In addition, case studies of ACI \citep{Li2023Large-EddyInteraction, Li2023ProcessAtlantic} may only represent a single realization of ACI for specific cloud regimes. Nevertheless, case studies are still useful for testing new physical processes that are important to case-dependent ACI metrics. We remark that the stochasticity of ACI discussed here is based on observations trained by the data-driven RFM instead of on first principles. Whether this is the true nature of ACI remains open.    

 Moving forward, one may study the stochasticity of ACI by applying data-driven algorithms to multi-field campaigns covering a wider range of different aerosol conditions and cloud regimes. Other than the stochasticity of $N_c$ prediction, one may explore the stochasticity of cloud macrophysical responses to aerosols (e.g., liquid water and cloud fraction adjustments to aerosol perturbations). 

\clearpage
\acknowledgments

This work was supported through the
ACTIVATE Earth Venture Suborbital-3 (EVS-3) investigation, which is
funded by NASA’s Earth Science Division and managed through the
Earth System Science Pathfinder Program Office.
C.V. is funded by DFG SPP-1294 HALO under project no 522359172 and by European Union under grant no 101101999 and by SESAR JU CICONIA.
The Pacific Northwest National Laboratory (PNNL) is
operated for the U.S. Department of Energy by Battelle Memorial Institute
under contract DE-AC05-76RLO1830.
The simulations were performed using resources available through
Research Computing at PNNL.

\section*{Availability Statement}

The source code used for the simulations of this study, the
Weather Research and Forecasting (WRF) model \citep{Li2023Xiang-yu/WRF-LASSO:WRF-LASSO},
is freely available on \url{https://zenodo.org/records/10421287}.
 The source code for the random forest model is publicly available at \url{https://scikit-learn.org/stable/modules/generated/sklearn.ensemble.RandomForestRegressor.html}. 
ACTIVATE observational data \citep{Team2020AerosolDataset} are publicly available at \url{https://asdc.larc.nasa.gov/project/ACTIVATE}.

\ifarXiv
\clearpage
\renewcommand{\thefigure}{A\arabic{figure}}
\renewcommand{\thetable}{A\arabic{table}}
\setcounter{figure}{0}
\setcounter{table}{0}
\section*{Appendices}
\begin{sidewaystable}
\caption{
  List of data and the corresponding instruments used in this study. n/a – not applicable.
} 
\centering
\setlength{\tabcolsep}{1pt}
\begin{tabular}{|c|c|c|c|}
\hline
  Instruments & quantities & resolution (uncertainty) & frequency (Hz) \\
  \hline
  FCDP & cloud droplet diameter & $3-50\, \mu{\rm m}$ ($\le 20\%$) & 1\\
  \hline
  2DS & cloud droplet diameter & $11.4-1464.9\, \mu{\rm m}$ ($\le 20\%$) & 1\\

  \hline
  Five-port pressure system (TAMMS) & $u, v, w$ & n/a($w: 10 \rm{cm\, s}^{-1}$; $u, v: 50 \rm{cm\, s}^{-1}$) & 20 \\
  \hline
  Rosemount 102 sensor & $T$ & n/a( $0.5 ^\circ$) & 1\\
  \hline
  Diode laser hygrometer (DLH) & $q_v$ & n/a (5\% or 0.1 ppmv) & $< 0.05$ \\
  \hline
  AMS & non-refractory aerosol mass concentration & 60-600 nm (50\%) & 1 \\
  \hline
  SMPS & aerosol particles with $d\le 100$ nm & $\le\pm10\%-20\%$ & 1/60 \\
  LAS & aerosol particles with $d\ge 100$ nm & $\le\pm10\%-20\%$ & 1 \\
\hline
\multicolumn{4}{p{0.3\textwidth}}{}
\end{tabular}
\label{tab:instrument}
%\end{table*}
\end{sidewaystable}

\begin{figure}[t!]\begin{center}
\begin{overpic}[width=\textwidth]{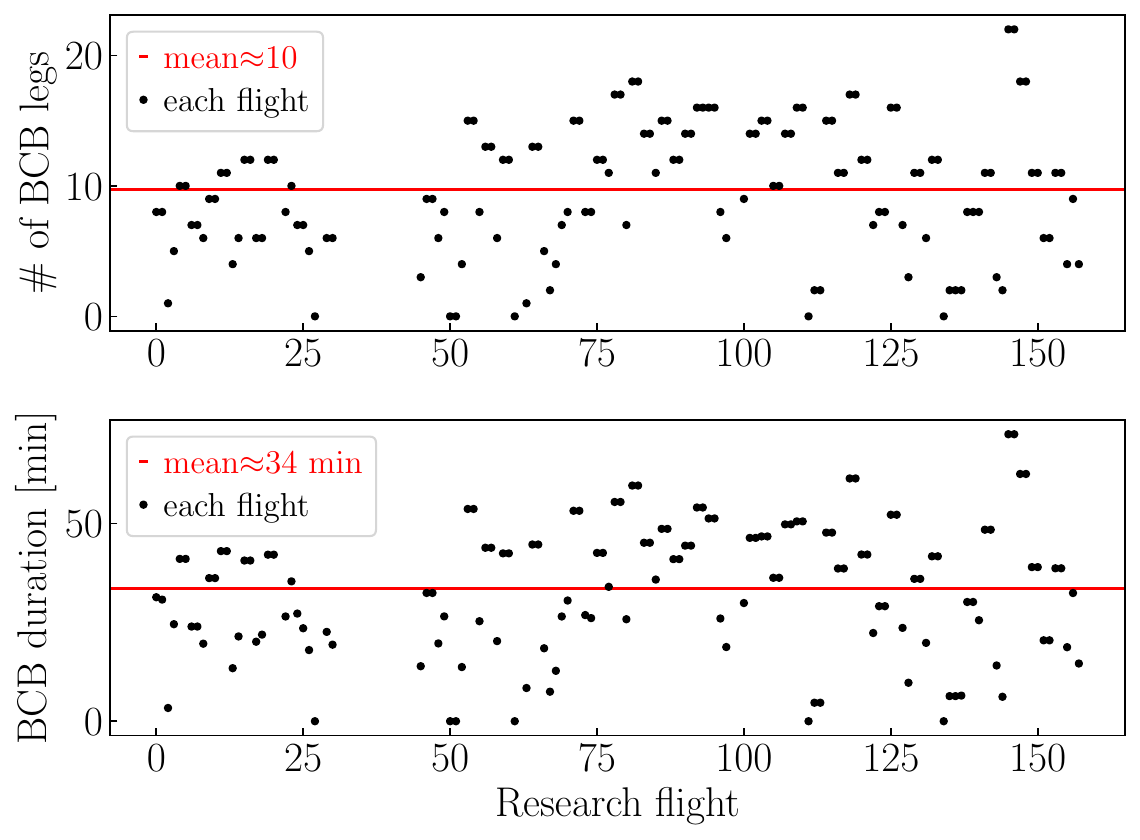}\end{overpic}
\end{center}\caption{Number and the total duration of BCB flight legs for each flight. The mean duration for each BCB leg of a flight is $34/10\approx3$ min. 
}
\label{num_duration_BCB}
\end{figure}

\begin{figure*}[t!]\begin{center}
\begin{overpic}[width=\textwidth]{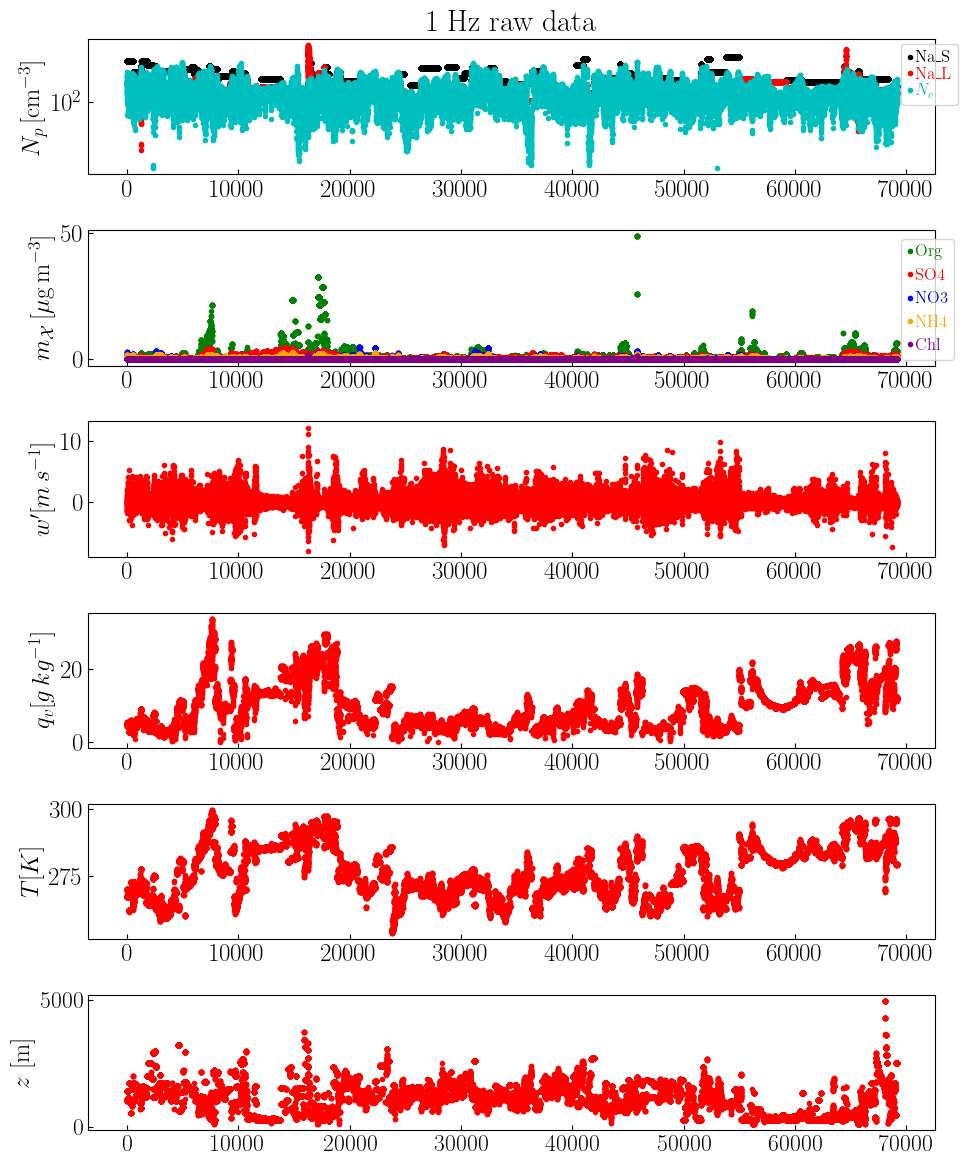}\end{overpic}
\end{center}\caption{1 Hz raw data of assumed physically important quantities.
The sample size is 69159 with a sampling rate of 1 Hz, which is from all the
ACTIVATE flights.
}
\label{activate_Na_X_wp_qv_T}
\end{figure*}

\begin{figure*}[t!]\begin{center}
\begin{overpic}[width=\textwidth]{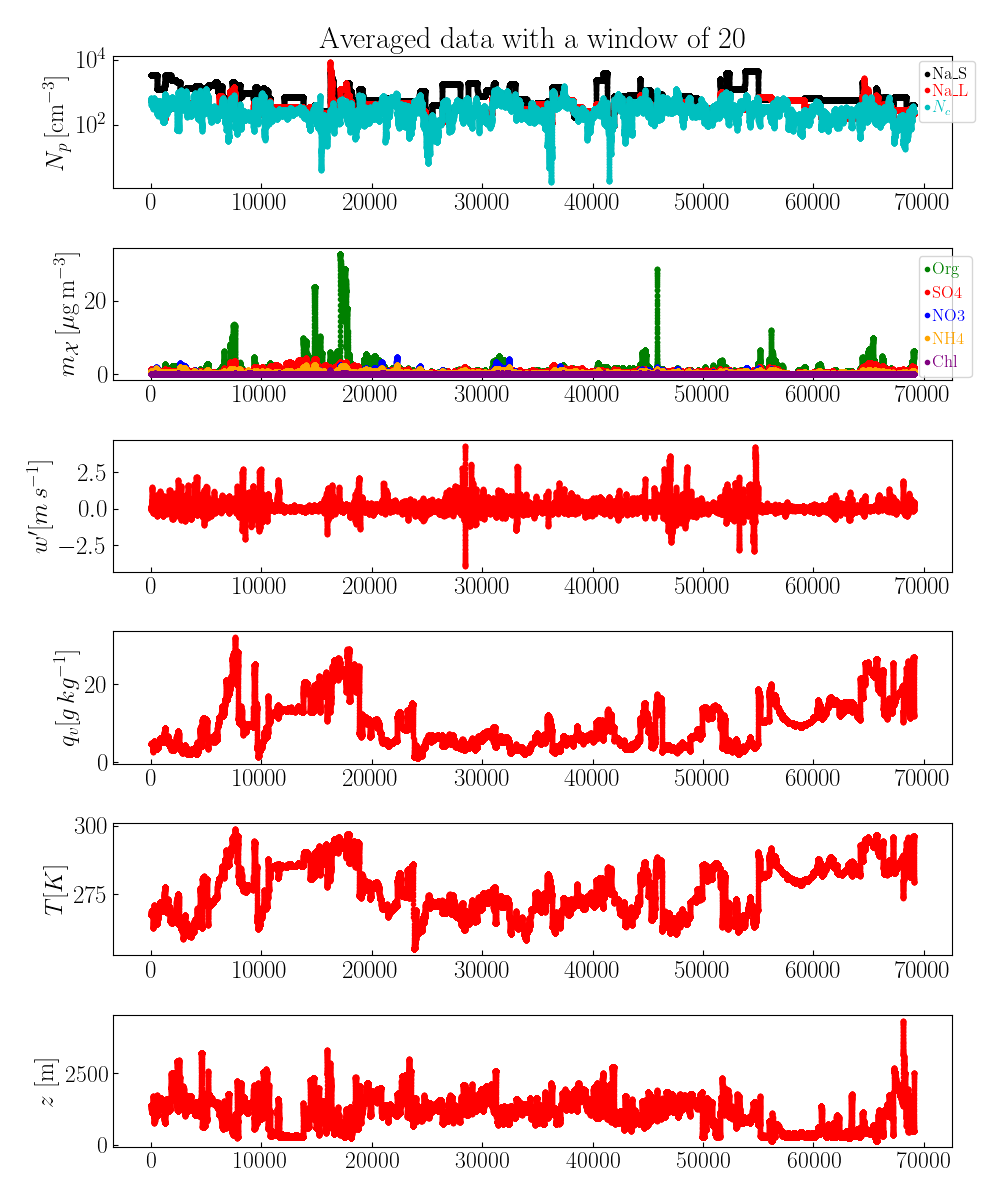}\end{overpic}
\end{center}\caption{Averaged data with a window size of 20 (data points) from the raw data
shown in \Fig{activate_Na_X_wp_qv_T} as the input dataset for the RFM algorithm.
}
\label{activate_Na_X_wp_qv_T_window20}
\end{figure*}

\def\fx{7}
\def\fy{85}
\begin{figure*}[t!]\begin{center}
\begin{overpic}[width=0.48\textwidth]{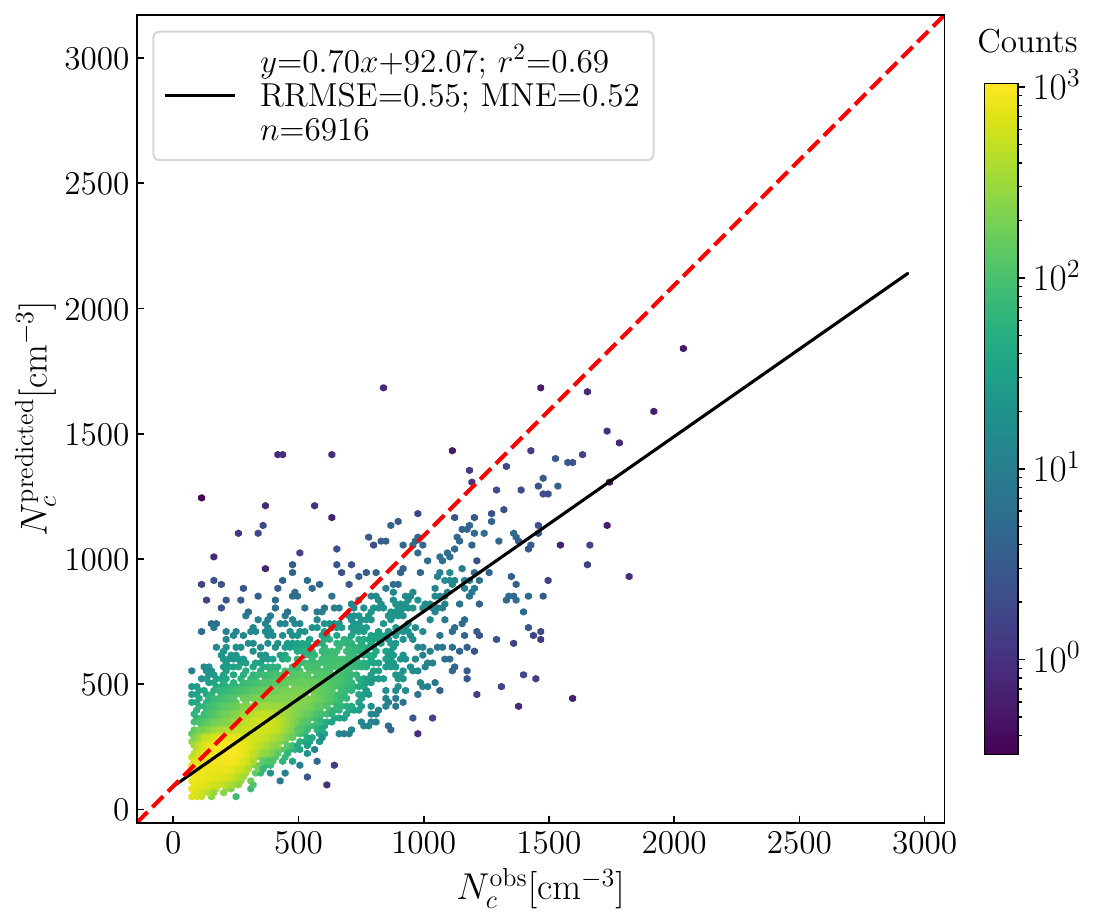}\put(\fx, \fy){1 Hz}\end{overpic}
\begin{overpic}[width=0.48\textwidth]{Nc_predicted_obs_20window_}\put(\fx, \fy){0.05 Hz}\end{overpic}
\end{center}\caption{Comparison between \nco and \ncp for different average window size as shown in \Tab{tab:windowSize} for the
full dataset ($N_c = \mathcal{G}(m_\mathcal{X}, N_a, w^\prime, u^\prime, v^\prime,
T, q_v, \bm{x}, \theta_z)$).
}
\label{fig:Nc_predicted_obs_20window}
\end{figure*}

\begin{table*}[t!]
\caption{
Scores for different average window size of data.
$N_c = \mathcal{G}(m_\mathcal{X}, N_a, w^\prime, u^\prime, v^\prime, T, q_v, \bm{x}, \theta_z)$.
Comparison between the prediction and observation is shown in \Fig{fig:Nc_predicted_obs_20window}.
} 
\centering
\setlength{\tabcolsep}{1pt}
\begin{tabular}{|c|c|c|c|c|}
\hline
 Window size & $r^2$ training score & $r^2$ validation score & OOB score \\
\hline
 %1 & 0.96 & 0.71  & 0.72 \\ 
 1 & 0.95 & 0.69  & 0.68 \\ 
% 2 & 0.97 & 0.81  & 0.82 \\ 
 2 & 0.97 & 0.81  & 0.80 \\ 
 5 & 0.99 & 0.93  & 0.93 \\ 
 10 & 1.00 & 0.98  & 0.98 \\ 
 20 & 1.00 & 0.99  & 0.99 \\ 
 50 & 1.00 & 1.00  & 1.00\\ 
\hline
\multicolumn{4}{p{0.3\textwidth}}{}
\end{tabular}
\label{tab:windowSize}
\end{table*}

\begin{figure}[t!]\begin{center}
\begin{overpic}[width=0.48\textwidth]{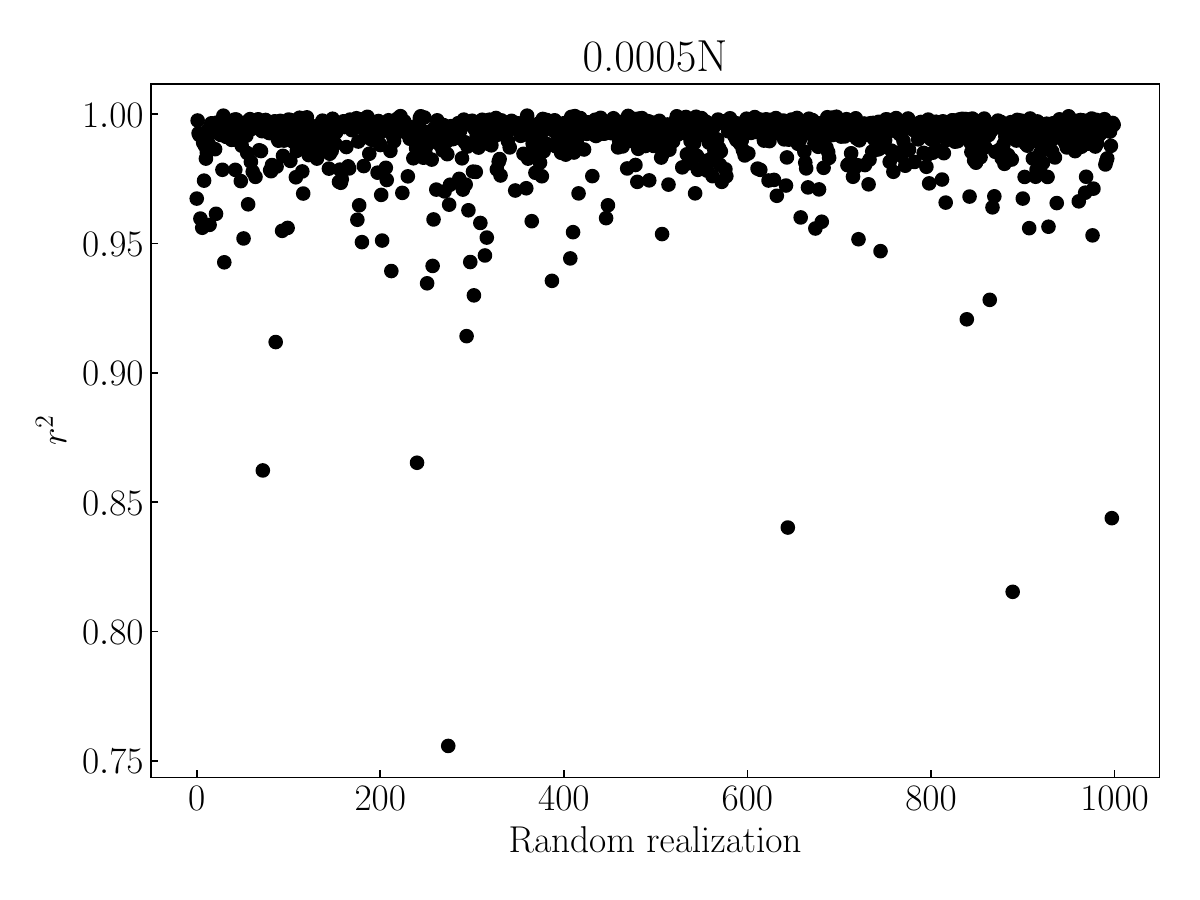}\end{overpic}
\end{center}
\caption{$r^2$ of 1000 random realizations for $0.0005N$ in \Fig{fig:Nc_predicted_obs_size}(a). Note that the samples are drawn from the averaged (window size 20) dataset.}
\label{fig:r2_size}
\end{figure}

\begin{figure*}[t!]\begin{center}
\begin{overpic}[width=0.32\textwidth]{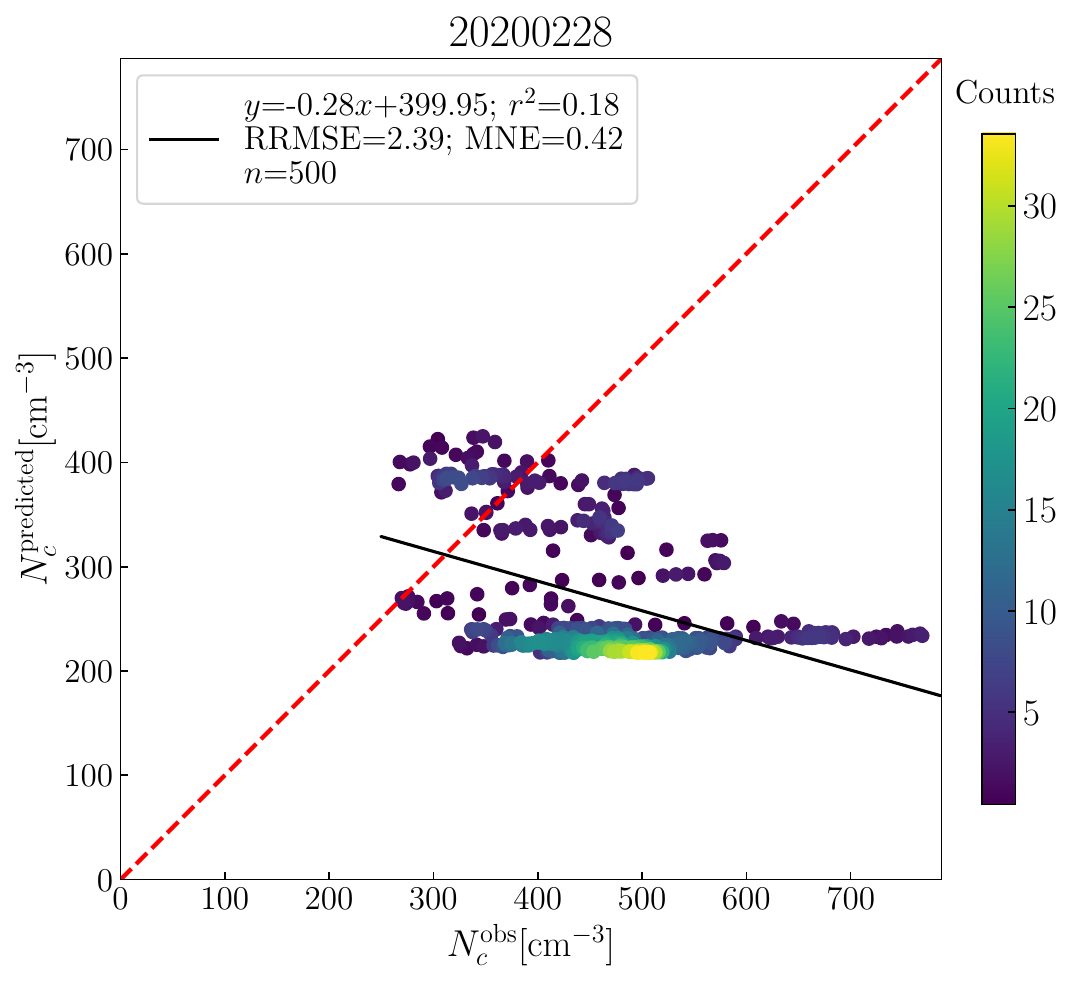}\end{overpic}
\begin{overpic}[width=0.32\textwidth]{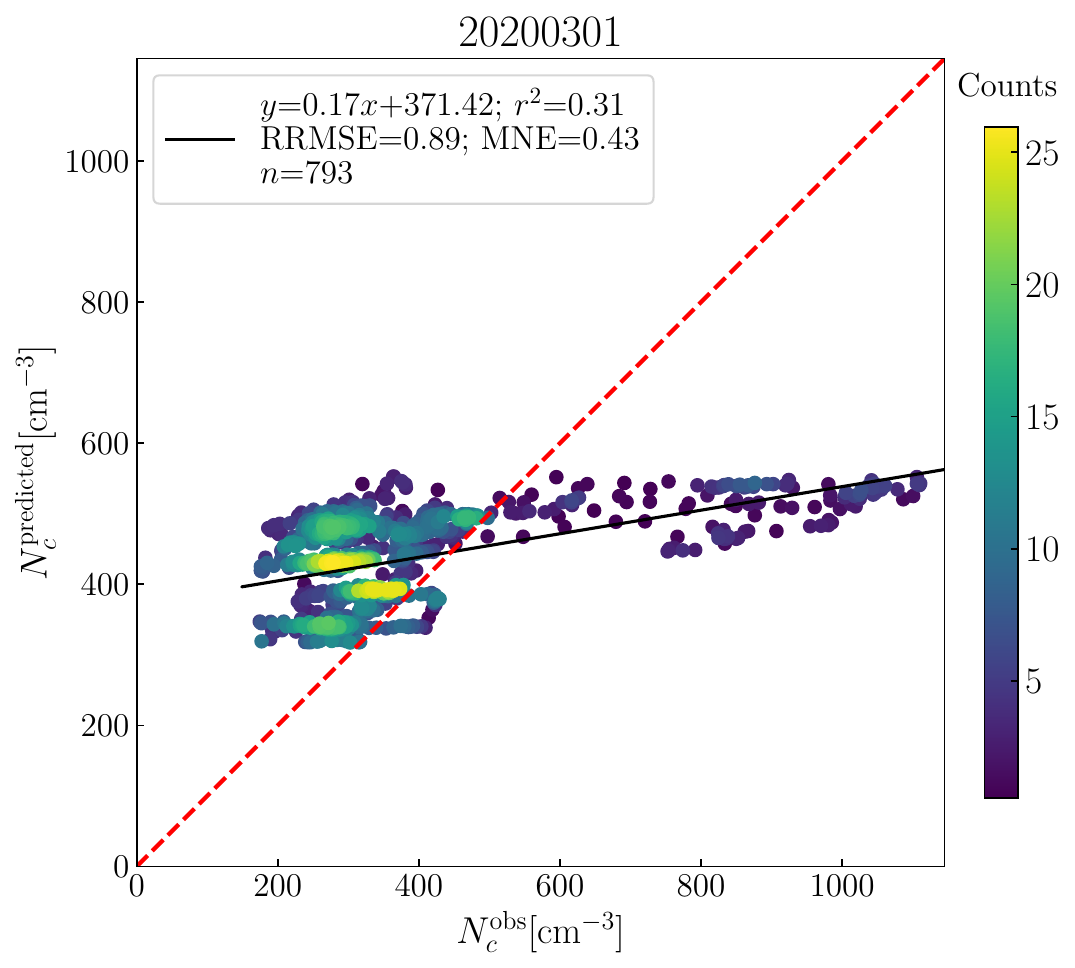}\end{overpic}
\begin{overpic}[width=0.32\textwidth]{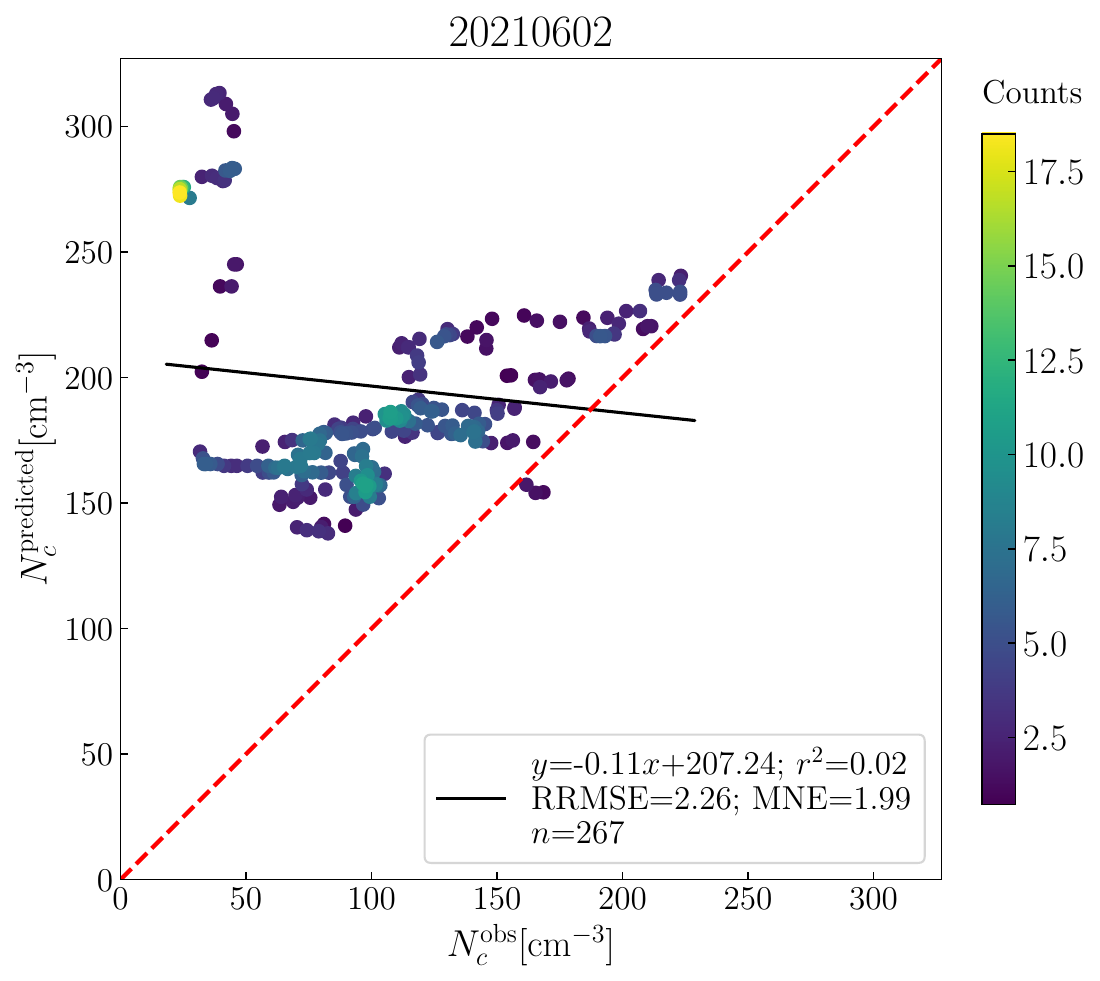}\end{overpic}
\begin{overpic}[width=0.32\textwidth]{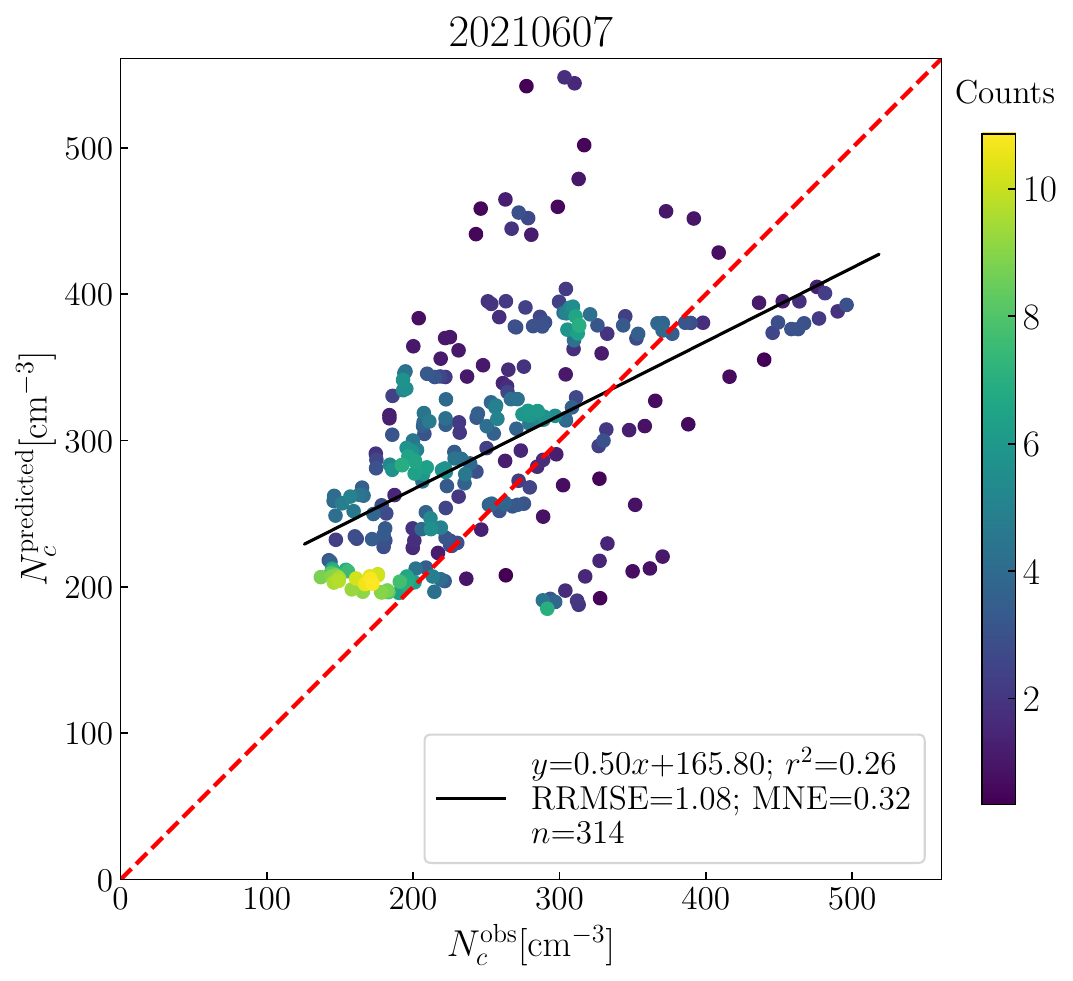}\end{overpic}
\begin{overpic}[width=0.32\textwidth]{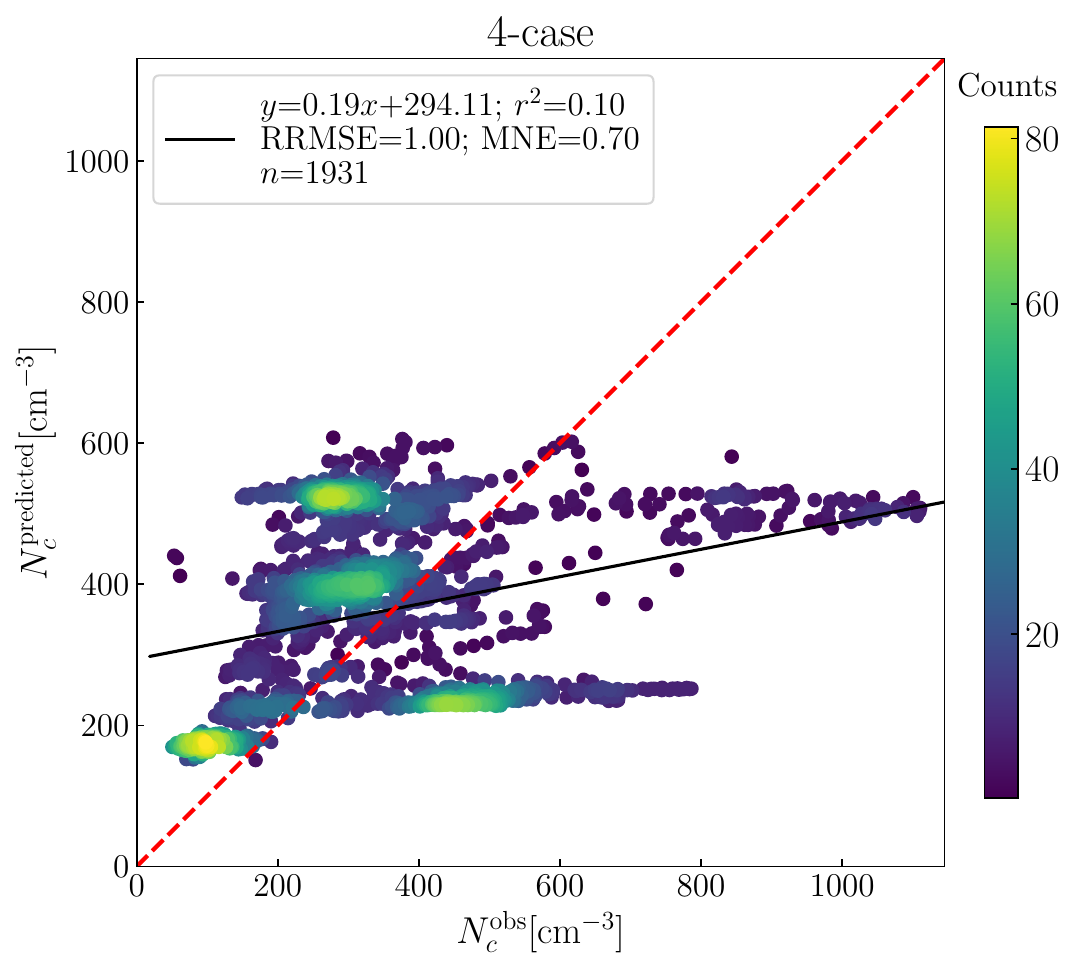}\end{overpic}
\end{center}
\caption{Same as \Fig{fig:Nc_predicted_obs_20200301} but with the same predictors as in \Fig{fig:Nc_predicted_obs}(a).}
\label{fig:Nc_predicted_obs_cases_allPredictors}
\end{figure*}

\begin{figure*}[t!]\begin{center}
\begin{overpic}[width=0.48\textwidth]{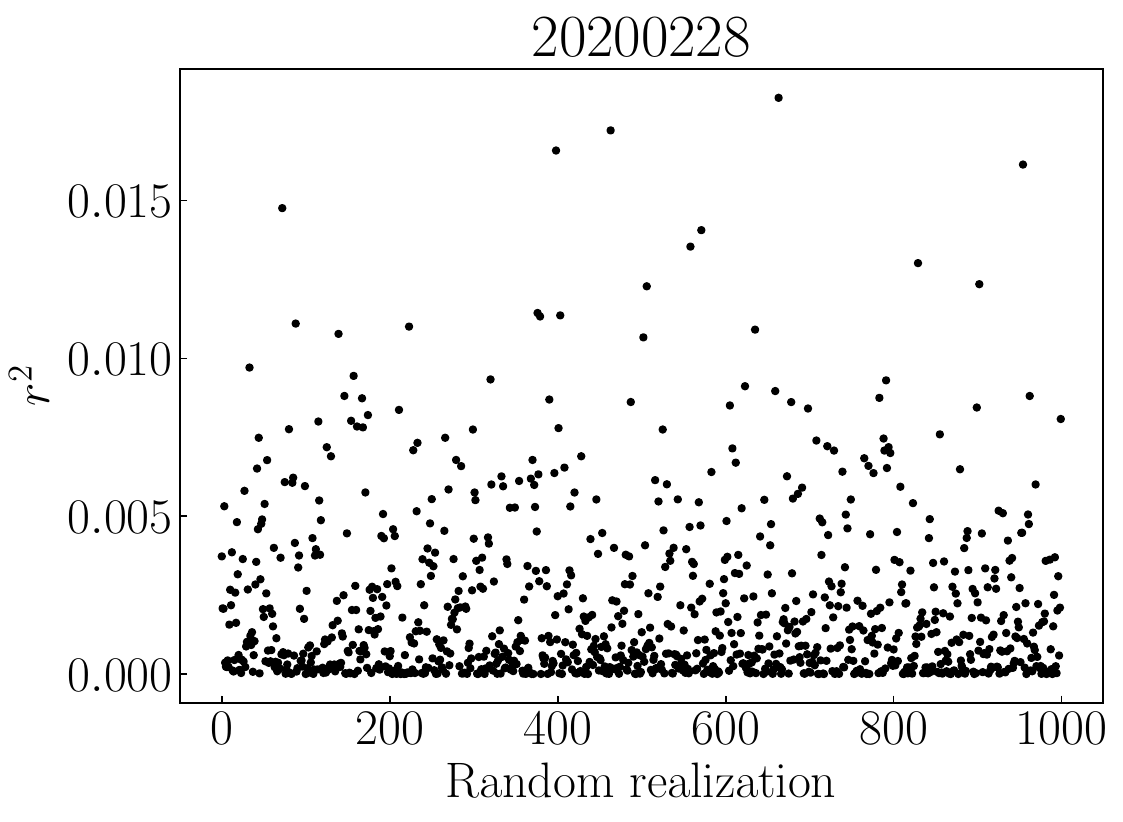}\end{overpic}
\begin{overpic}[width=0.48\textwidth]{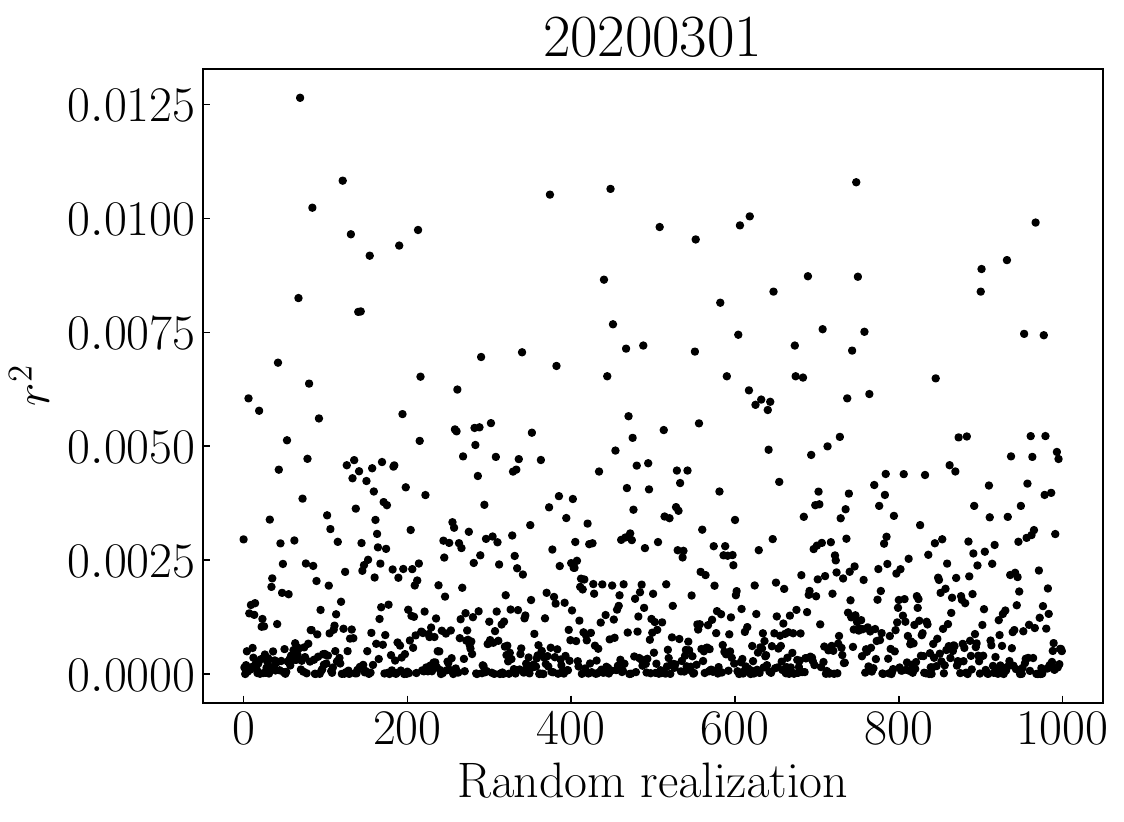}\end{overpic}
\begin{overpic}[width=0.48\textwidth]{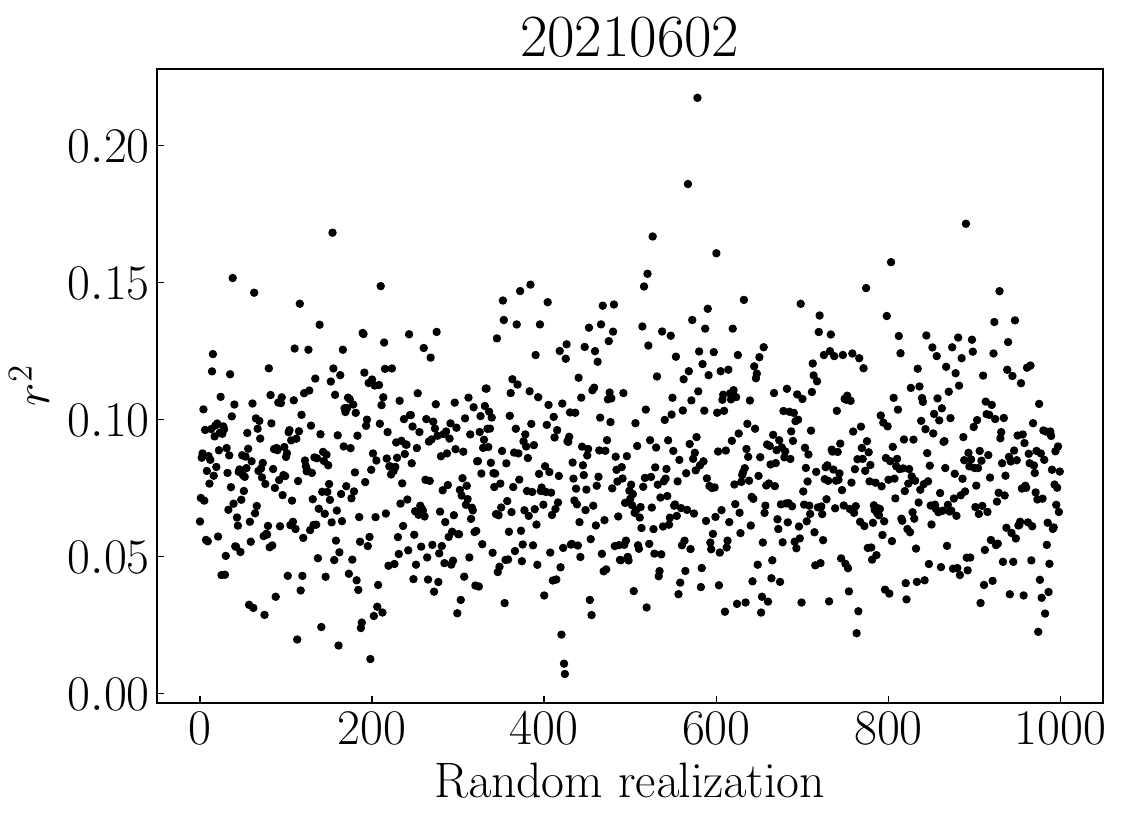}\end{overpic}
\begin{overpic}[width=0.48\textwidth]{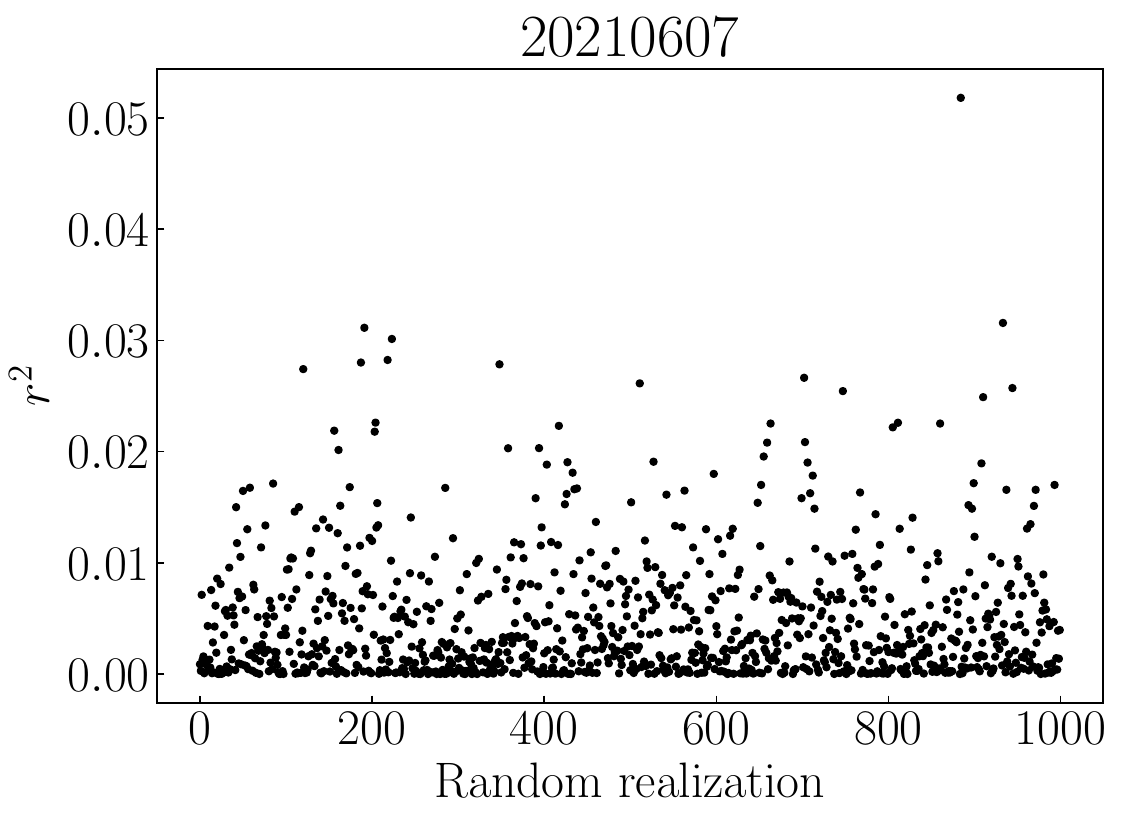}\end{overpic}
\end{center}
\caption{$r^2$ between the FCDP measurements and LES. As the LES data were sampled randomly at each height, we test how the random selection affects the $r^2$ by performing 1000 realizations for each case.}
\label{fig:r2_FCDP_LES}
\end{figure*}

\begin{figure*}[t!]\begin{center}
 \begin{subfigure}[b]{0.49\textwidth}
      \subcaption{}
      \includegraphics[width=\textwidth]{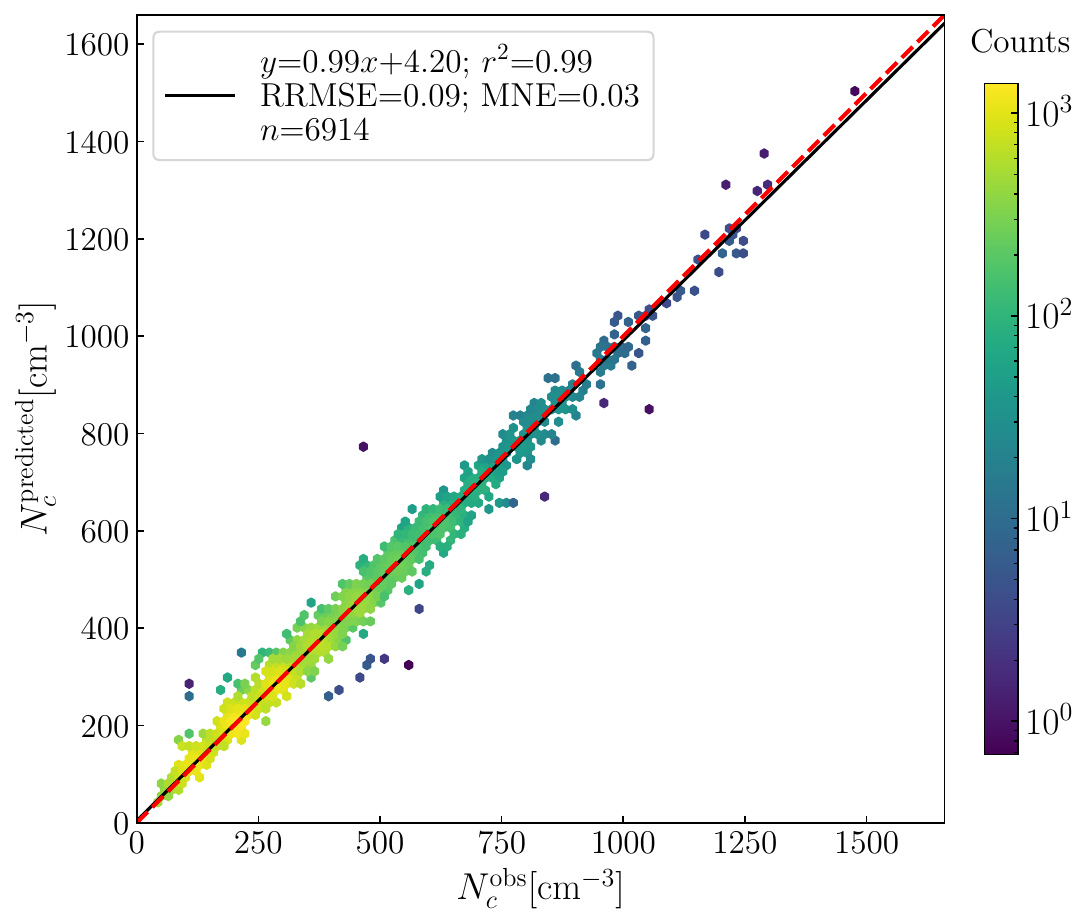}
  \end{subfigure}
 \begin{subfigure}[b]{0.49\textwidth}
      \subcaption{}
      \includegraphics[width=\textwidth]{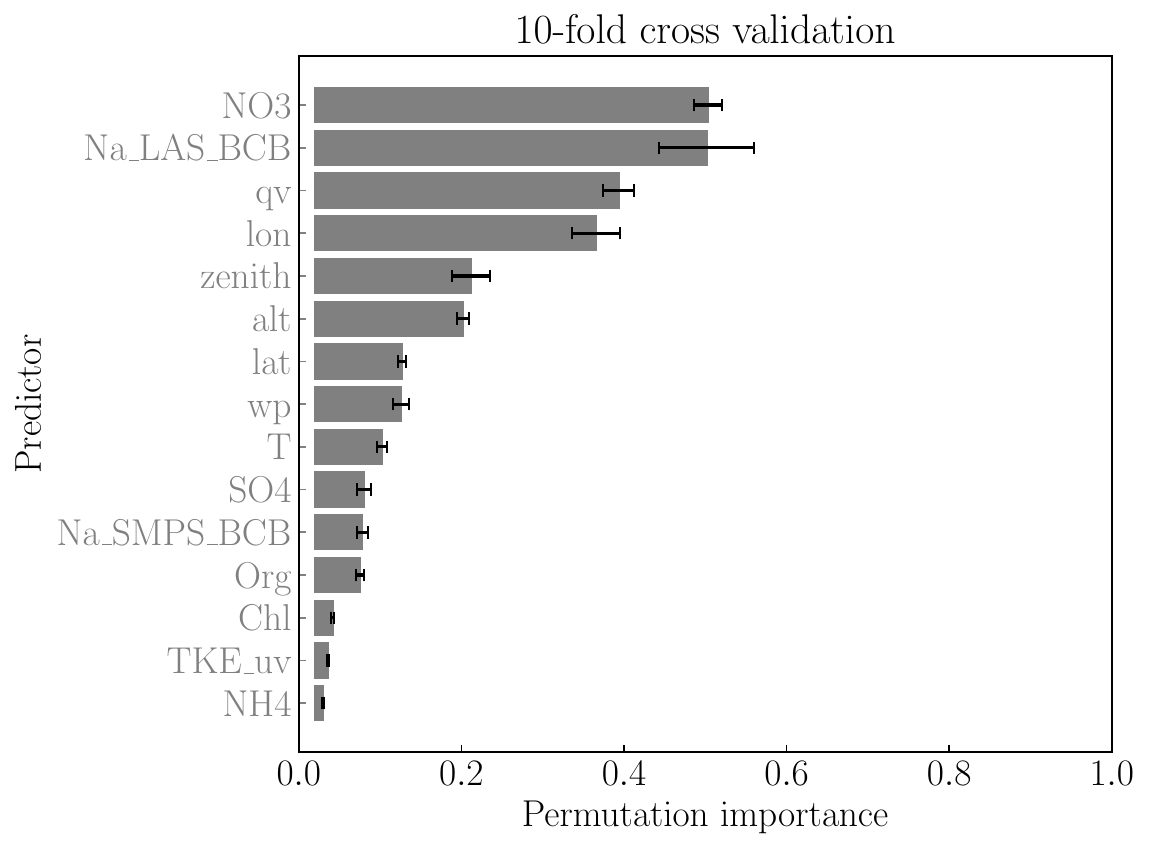}
  \end{subfigure}
%\begin{overpic}[width=0.48\textwidth]{Nc_predicted_obs_20window_TKE_uv}\end{overpic}
% \begin{overpic}[width=0.48\textwidth]{K_folds20window_TKE_uv}\end{overpic}
\end{center}\caption{Same as \Fig{fig:Nc_predicted_obs}(b) but with $N_c = \mathcal{G}(m_\mathcal{X}, N_a, w^\prime, \text{TKE}\_{uv},
T, q_v, \bm{x}, \theta_z)$.
}
\label{fig:Nc_predicted_obs_tke}
\end{figure*}

\fi

\end{document}